%% file: SemSpace2020JournalVersionExp.tex
\documentclass{jcs}
\usepackage{graphicx}

\input{tikzpreamble.tex}

\usepackage{prftree, proof, stmaryrd}
\usepackage{mathtools, float, amstext, cancel}
\usepackage{hyperref, enumerate, array}
\usepackage[all,cmtip]{xy}

\usepackage{booktabs,subcaption}

\usepackage{amsmath}
\newtheorem{definition}{Definition}

\newcommand{\blstar}{\mathbf{!L^*}}
\newcommand{\LA}{\mathbf{LA}}

\newcommand{\Vect}{\mathbf{Vect}}

\newcommand{\ev}{\mathrm{ev}}
\newcommand{\Typ}{\mathrm{Typ}}

\newcommand{\F}{\mathcal{F}}

\newcommand{\R}{\mathbb{R}}
\newcommand{\fdVect}{\mathbf{FdVect}}

\newcommand{\bs}{\backslash}
\renewcommand{\L}{\mathbf{L}}

\newcommand{\ov}[1]{\overrightarrow{#1}}

\renewcommand{\epsilon}{\varepsilon}

\newcommand{\semantics}[1]{\llbracket #1 \rrbracket } 

\begin{document}
\title{\bf Anaphora and Ellipsis in Lambek Calculus with a Relevant Modality: Syntax and Semantics
\date{}
}
\titlerunning{Anaphora and Ellipsis in Lambek Calculus with a Relevant Modality}
\authorrunning{McPheat et al.}

\author[1]{Lachlan McPheat}
\author[2]{Gijs Wijnholds}
\author[1]{Mehrnoosh Sadrzadeh}
\author[2]{Adriana Correia}
\author[3]{Alexis Toumi}

\affil[1]{University College London, UK} 
\affil[2]{Utrecht University, NL}
\affil[3]{Oxford University, UK}


\maketitle

\begin{abstract}
     Lambek calculus with a relevant modality $\blstar$ of  \citep{Kanovich2016} syntactically resolves parasitic gaps in natural language. It resembles the Lambek calculus with anaphora $\LA$ of  \citep{jager1998multi} and the Lambek calculus with controlled contraction $\mathbf{L}_{\Diamond}$ of  \citep{WijnSadr2019} which deal with anaphora and ellipsis. What all these calculi add to Lambek calculus is a  copying and moving behaviour.  Distributional semantics is a subfield of Natural Language Processing that uses vector space semantics for words via  co-occurrence statistics in  large corpora of data. Compositional vector space semantics for Lambek Calculi are obtained via the \emph{DisCoCat} models \citep{Coeckeetal2010}.  $\LA$ does not have a vector space semantics and the semantics of $\mathbf{L}_{\Diamond}$ is not compositional. Previously, we  developed a DisCoCat semantics for $\blstar$ and focused on the parasitic gap applications. In this paper, we use the vector space instance of that general semantics  and show how one can also  interpret anaphora, ellipsis, and for the first time derive the sloppy vs strict vector readings of ambiguous anaphora with ellipsis cases. The base of our semantics is tensor algebras and their finite dimensional variants: the Fermionic Fock spaces of Quantum Mechanics. We implement our model and experiment with the ellipsis disambiguation task of  \citep{WijnSadrNAACL}. 

\end{abstract}


\section{Introduction}
	Inspired by Linear Logic's $!$, extensions of Lambek calculi with exponentials were introduced in the 90's in \citep{Morrilletal1990,Barryetal1995} for the purpose of deriving phenomena with medial extraction, as in relative clauses such as ``\textit{book which fell}'', ``\textit{book which John read}'', and ``\textit{book which John read yesterday}'',  iterated coordination, as in ``\textit{John Bill Mary and Suzy}'', parasitic gaps, as in ``\textit{the paper which Suzy filed without reading}'', and  cases were there is a lexical item whose arguments optionally contains a gap, as in the lexical item `\textit{too}' in clauses such as ``\textit{too boring for one to follow}''. For a modern exposition and review of these systems, see \citep{MootRetore2012}. Similar extensions, with limited copying, and combinations of permutation and associativity, rather than just permutation,  were used to model coreference phenomena such as anaphora and ellipsis in \citep{jager1998multi,morrill1996generalising}. These systems were recently revisited in \citep{morrill2016logic,morrill2015computational} and also independently in \citep{WijnSadr2018,WijnSadr2019}. It was the former set that inspired the logic $\blstar$: Lambek calculus with a Relevant Modality, developed in \citep{Kanovich2016} and used in  this paper. 

	In previous work \citep{mcpheat2020categorical}, we developed a categorical semantics for $\blstar$ via coalgebra modalities of Differential Categories \citep{St2006} and showed how at least two different interpretations of $!$ in the category of finite dimensional vector spaces and linear maps provide vector space instantiations for the abstract categorical semantics. Our focus in that paper, was to develop a vector space semantics for parasitic gaps, the motivating example of \citep{Kanovich2016}. In this paper, we briefly re-introduce the vector space semantics of $\blstar$, making use of the notion of tensor algebras and their finite variants: Fermionic Fock spaces of Quantum Mechanics. Then, following previous work \citep{WijnSadr2018,WijnSadr2019} and the work of J\"{a}ger \citep{jager1998multi,jager2006anaphora}, we apply the setting to obtain tensor algebraic interpretations of examples with anaphora, ellipsis, and combinations of the two. By so doing, we overcome two different weaknesses of the above mentioned previous approaches: (1) developing a compositional vector space semantics for a Lambek calculus that can deal with anaphora and ellipsis, whereas the work of \citep{jager1998multi,jager2006anaphora} did not (it only had a lambda calculus semantics), and (2) being able to obtain two different vector  semantics for the sloppy vs strict readings of ambiguous  anaphora with ellipsis cases, whereas the work in  \citep{WijnSadr2018,WijnSadr2019} could not.  The contributions of this  and previous work  \citep{mcpheat2020categorical} are thus two-fold: first we have developed a direct vector space semantics for an extension of Lambek calculus with limited contraction. The work done in  \citep{WijnSadr2018,WijnSadr2019} does develop a vector space semantics for J\"{a}ger's multi modal calculus \citep{jager1998multi}, but via a two-step semantics and by going through a dynamic logic of vectors \citep{muskens2019context}. Turning  this semantics into a single-step one faced the challenge of  discovering a  linear copying map. We have done this for $\blstar$ through the use of Fermionic tensor algebras and in effect infinitely many linear copying maps. Second, the semantics proposed in  \citep{WijnSadr2018,WijnSadr2019}, could not distinguish  between the sloppy and strict readings of anaphora with ellipsis, and here we are able to do so.  We implemented our vector space semantics on  distributional data and experimented with the ellipsis disambiguation dataset of  \citep{WijnSadrNAACL}. The results show that our linear copying operations do as well as the full non linear copying and also outperform non compositional and additive baselines. 

	Finally, the use of exponentials may not be  necessary when dealing with the phenomena of  parasitic gaps, anaphora and ellipsis; milder modal Lambek calculi with structural control,  as defined in \citep{moortgat1996multimodal},  or displacement calculi \citep{morrill2011displacement}, can also be used as a base as well, e.g. see the recent proposal of \citep{Moortgatetal2019}, which uses modal Lambek calculi with a Frobenius algebraic semantics for parasitic gaps. Finding  formal connections with these is a possible future direction. 

\section{Background}
	We introduce the relevant background and context for this paper, along with the definitions and examples of the linguistic phenomena we are considering (i.e. anaphora and ellipsis), the logic we employ, and previous examples of similar logics modelling anaphora and ellipsis.
	\subsection{Anaphora and Ellipsis}\label{subsec:anaphoraAndEllipsis}
		Anaphora and Ellipsis are instances of \textbf{coreference} in natural language, which is a phenomenon where two distinct linguistic expressions refer to the same entity \citep{ChomskyGovernment,KentEllipsis}. Resolving coreference is the task of identifying what entity is being referred to. 
		
		\textbf{Anaphora} is a phenomenon where  expressions called \emph{anaphors}  receive their meaning from a previously mentioned word or phrase, for instance pronouns like  `\textit{He}'  in `\textit{John sleeps. He snores.}', where clearly, `\textit{John}' and `\textit{He}' have the same referent\footnote{One may also consider \textbf{cataphora}, which are phenomena  where  expressions called \emph{cataphors} receive their meanings from a word or phrase following them.}.
		
		\textbf{Ellipsis} is a more general phenomenon, where broadly speaking, the meanings of entire phrases are referred to by combinations of other words,  sometimes called \emph{ellipsis markers}. In some cases ellipsis markers are dropped and a phrase is referred to by the empty expression. For instance, in  ``\textit{John plays guitar. Mary does too}.", the combination of words  `\textit{does too}'  is an ellipsis marker that refers to the phrase  `\textit{plays guitar}'.  A complete list of types of ellipsis is not yet agreed upon in the linguistics community and different theories have different lists. The above example is sometimes called \textit{verb-phrase ellipsis} (VP-ellipsis) and is the focus of this paper.
		
		\textbf{Anaphora with ellipsis} is  an utterance containing \textit{both} anaphora and ellipsis. The example we will use throughout this paper is ``\textit{John likes his code. Bill does too}". Here the pronoun `\textit{his}' is an anaphor, and the verb phrase `\textit{likes his code}' is being elliptically referred to by  the marker `\textit{does too}'. Note  how we can read this example in two different ways by choosing which order to resolve the anaphora or the ellipsis. The simpler way is called the \textbf{strict} reading, and has the meaning ``\textit{John likes John's code. Bill likes John's code}", which is achieved by first resolving the anaphor with \textit{John}, and then the ellipsis with the verb-phrase `\textit{likes John's code}'. The second reading is called the \textbf{sloppy} reading, and is achieved by first resolving the ellipsis with `\textit{likes his code}', and then resolve the  two distinct anaphors  `\textit{his}'   with `\textit{John}' and `\textit{Bill}' separately, giving us the meaning ``\textit{John likes John's code. Bill likes Bill's code}". In section \ref{sec:coreferenceInBlstar} we show how to distinguish between the two readings syntactically, and in section \ref{sec:VSScoreference} we derive the vector semantics of the two readings.
		
	\subsection{Type Logical Grammars and Anaphora}\label{sec:TLGAnaphora}
			Earlier works have modelled the syntactic structures that arise from anaphora and ellipsis as derivations in type-logical grammars (TLGs) in efforts to model these phenomena in the same formalism as TLGs model other phenomena of language.
			An early instance of this work comes in a paper by J\"{a}ger \citep{jager1998multi}, where the author defines a multimodal TLG which syntactically resolves anaphora and ellipsis.
			 J\"{a}ger follows up this idea in his book \citep{jager2006anaphora}, where he suggests the more concise idea of using TLGs with controlled contraction and permutation. However, neither in this book nor in \citep{jager1998multi} can you find any mention of a  vector space semantics, as this was not the goal of the author. A standard lambda calculus semantics was however developed. 

			  A vector space semantics for VP-ellipsis is developed in \citep{WijnSadr2018}, which employs a TLG with controlled contraction and permutation, inspired by \citep{jager2006anaphora}, and defines  vector space interpretations for anaphora with ellipsis examples. This semantics, however, uses a nonlinear copying operation, and fails to distinguish between strict and sloppy readings. Both of the issues with this semantics are addressed in this paper.

	\subsection{Lambek Calculus with a Relevant Modality}
		Lambek calculus with a Relevant Modality, denoted by $\blstar$,  is a variation of  Lambek calculus, $\L$ where we may use empty sequents (denoted by $*$), and which has a (relevant) modality, $!$, which denotes when a formula is not resource sensitive. We consider the calculus $\blstar$ over the set of primitive types $\{N,S\}$ and logical connectives $\{/,\bs, !\}$. The primitive types represent the basic grammatical types of  ``noun" and ``sentence". With these types and the connectives we generate the full set of types\footnote{
		We use the terms `formula' and `type' interchangeably.} according to the following BNF: 
	\[
    	\text{Typ}_{\blstar} ::= * \mid N\mid S\mid A,B \mid A/B \mid A\bs B \mid !A.
	\]
		The rules of $\blstar$ are presented in the sequent calculus system of  table \ref{tab:syntaxRules}. Sequents are written as $\Gamma \to A$, where $\Gamma$ is a finite (possibly empty)  list of formulas $\{A_1,\ldots,A_n\}$, and $A$ is a single  formula. We refer to the set of formulas of $\blstar$ as $\Typ_\blstar$.
		
		\begin{table}[t!]
		\centering
		\scalebox{0.9}{\begin{tabular}{rrl}
		 
		 \prftree{A \to A}  
		 & 
		 \prftree[r]{$\scriptstyle{(/ L)}$}
		 {\Gamma \to A}
		 {\Delta_1, B, \Delta_2 \to C}
		 {\Delta_1, B / A, \Gamma, \Delta_2 \to C}
		 &
		 \prftree[r]{$\scriptstyle{(/ R)}$}
		 {\Gamma, A \to B}
		 {\Gamma \to B/A} 
		 \\ 
		 \\
		 &
		 \prftree[r]{$\scriptstyle{(\bs L)}$}
		 {\Gamma \to A}
		 {\Delta_1, B, \Delta_2 \to C}
		 {\Delta_1, \Gamma, A\bs B, \Delta_2 \to C} 
		 &
		 \prftree[r]{$\scriptstyle{(\bs R)}$}
		 {A, \Gamma \to B}
		 {\Gamma \to A \bs B}
		 \\
		 \\
		 \prftree[r]{$\scriptstyle{(!L)}$}
		 {\Gamma_1, A, \Gamma_2 \to C}
		 {\Gamma_1, !A, \Gamma_2\to C} 
		 &
		 \prftree[r]{$\scriptstyle{(!R)}$}
		 {!A_1,\ldots, !A_n \to B}
		 {!A_1,\ldots, !A_n \to !B}
		 \\
		 \\
		  \prftree[r]{$\scriptstyle{(perm_1)}$}{\Delta_1, !A,\Gamma, \Delta_2\to C}{\Delta_1, \Gamma, !A, \Delta_2 \to C}  
		  &
		 \prftree[r]{$\scriptstyle{(perm_2)}$}{\Delta_1, \Gamma, !A, \Delta_2 \to C}{\Delta_1, !A,\Gamma, \Delta_2\to C}  
		 &
		 \prftree[r]{$\scriptstyle{(contr)}$}
		 {\Delta_1, !A, !A, \Delta_2 \to C}
		 {\Delta_1, !A, \Delta_2 \to C}
		 \\
		\end{tabular}}
		\caption{Rules of the $\blstar$ calculus.}\label{tab:syntaxRules}
		\vspace{-0.7cm}
		\end{table}

\section{Resolving Anaphora and Ellipsis in $\blstar$}\label{sec:coreferenceInBlstar}
	We follow \citep{WijnSadr2018,jager1998multi,jager2006anaphora}  in how we resolve anaphora and ellipsis syntactically using type logical grammars.
	Given a word or compound $w$ with a   $\blstar$-type $C$ whose meaning comes from a reference $w'$ (necessarily of the same type) e.g. a pronoun (anaphor) or an ellipsis marker (eg. `\textit{does-too}'), we first change the assignments of types of $w$ to $(!C\bs C)$. We then assign the reference word, $w'$ the type $!C$, apply contraction to this type, obtain  the formula $!C,!C$, and then move one of the copies to the reference type ($!C\bs C$), to which we then apply the $(\bs L)$-rule. This procedure provides us with a proof tree  as follows:
	\[
	\infer[(contr)]
		{\Gamma_1, !C, \Gamma_2, !C\bs C, \Gamma_3 \longrightarrow B}
		{
		\infer[(perm_2)]
			{\Gamma_1, !C, !C, \Gamma_2, !C\bs C, \Gamma_3 \longrightarrow B}
			{
			\infer[(\bs L)]
				{\Gamma_1, !C, \Gamma_2, !C, !C\bs C, \Gamma_3 \longrightarrow B}
				{\infer[]{!C \longrightarrow !C}{}
				&
				\infer[]{\Gamma_1, !C, \Gamma_2, C, \Gamma_3 \longrightarrow B}{\vdots}}}}
	\]
	where $\Gamma_1,\Gamma_2, \Gamma_3$ are (possibly empty) contexts surrounding the coreference.

	For a concrete example of this procedure, consider: ``\textit{John sleeps. He snores}". Clearly,  `\textit{He}' refers to `\textit{John}', but since`\textit{John}' is already being used as the argument of `\textit{sleeps}', it is not possible for `\textit{He}'   to use it as well in classical Lambek calculus. Thus we instead wish to ``copy" the meaning of `\textit{John}' using the $(contr)$-rule of $\blstar$, and move one copy to where `\textit{He}' can use it,  using the $(perm_2)$-rule of $\blstar$.

	In what follows, we present examples of  anaphora, ellipsis, and the combination of the two. These examples involve some seemingly intimidating derivations, of which we want the reader to be wary. We will verbally describe what is happening in each of these derivations, and later in section \ref{sec:VSScoreference} we will be able to view the derivations as string diagrams, which are far more intuitive.	
	\subsection{Anaphora}
		Following \citep{WijnSadr2018}, we work with the example ``\textit{John sleeps. He snores}" and the following type assignment
		\footnote{
		Note that one can always ``ignore'' the $!$ of a type by applying the $(!L)$-rule. This means that in principle one may wish to consider every noun and every verb phrase as having type $!$, since every noun/VP may be referred to at some point. We choose to exclude this typing, as it adds more steps to the derivations.}
		: 
		\[
		\{ (\mbox{John}, !N), (\mbox{sleeps}, N\bs S),(\mbox{snores}, N\bs S), (\mbox{he}, !N \bs N) \}
		\]
		 which provides us with the derivation proof tree in figure  \ref{fig:anaphora}.
 	\begin{figure}

 %
		\small
		 \[\scalebox{0.7}{
		 \infer[(contr)]
		 	{!N, N \bs S, !N \bs N, N\bs S \to S,S}
			{\infer[(perm_2)]
				{!N, !N, N \bs S, !N \bs N, N\bs S \to S,S}
				{\infer[(!L)]
				{!N, N \bs S, !N, !N \bs N, N\bs S \to S,S}
					{\infer[(\bs L)]
				{N, N \bs S, !N, !N \bs N, N\bs S \to S,S}
					{\infer[ ]{N \to N}{} 
					&
					\infer[(\bs L)]
					{S, !N, !N \bs N, N \bs S \to S,S}
						{\infer[]
						{!N \to !N}{}
						& 
						\infer[(\bs L)]{S,N,N \bs S \to S,S}{\infer[ ]{N \to N}{} 
							& \infer[]{S, S \to S,S}{}}}}}}}}\]
		\normalsize
		\caption{Anaphora Derivation}\label{fig:anaphora}
		\end{figure}
		We note that the very first rule applied (reading the proof from bottom to top) is the contraction which copies the type $!N$  of `\textit{John}' providing  us with $!N,!N$. We then permute the rightmost copy  to the immediate left hand side of the  type of  `\textit{He}', i.e. the formula  $!N\bs N$, and then unify  the types of `\textit{John}' and `\textit{He}' using the $(\bs L)$-rule.
		
	\subsection{Ellipsis}
		For ellipsis, consider the simple example ``\textit{John plays guitar. Mary does too}" and the following type assignment
		\small \[
		\left \{ (\mbox{John}, N), (\mbox{plays}, !(N\bs S)/N), (\mbox{guitar}, N), (\mbox{Mary}, N), (\mbox{does-too}, (!(N \bs S)) \bs (N \bs S))\right \}.
		\]\normalsize
		In order to  resolve the ellipsis in this example, all we need to do is to show that the sequent
		\[
		N, !(N\bs S)/N, N, N, (!(N\bs S))\bs (N\bs S) \longrightarrow S,S
		\]
		 is derivable in $\blstar$. This is done in figure \ref{fig:ellipsis}.
		\begin{figure}
		\[\scalebox{0.7}{
			\infer[(/L)]{N, !(N\bs S)/N, N, N, (!(N\bs S))\bs (N\bs S) \longrightarrow S,S}
				{\infer[]{N \longrightarrow N}{}
				&
				\infer[(contr)]{N, !(N\bs S), N, (!(N\bs S))\bs (N\bs S) \longrightarrow S,S}
					{\infer[(perm_2)]{N, !(N\bs S), !(N\bs S), N, (!(N\bs S))\bs (N\bs S) \longrightarrow S,S}
						{\infer[(\bs L)]{N, !(N\bs S), N,!(N\bs S), (!(N\bs S))\bs (N\bs S) \longrightarrow S,S}
							{\infer[]{!(N\bs S) \longrightarrow!(N\bs S)}{}
							&
							\infer[(!L)]{N, !(N\bs S), N, N\bs S \longrightarrow S,S}
								{\infer[(\bs L)]{N, N\bs S, N, N\bs S \longrightarrow S,S}
									{\infer[]{N \longrightarrow N}{}
									&
									\infer[(\bs L)]{S, N, N\bs S \longrightarrow S,S}
										{\infer[]{N \longrightarrow N}{}
										&
										\infer[]{S,S \longrightarrow S,S}{}}}}}}}}
		}\]
		\caption{Ellipsis Derivation}\label{fig:ellipsis}
		\end{figure}
	The first application of a rule is the ($/ L$)-rule, which applies `\textit{plays}' to its subject `\textit{guitar}'. After that application we proceed as we do in the anaphora example, the next rule being contraction, but this time to the formula of the entire verb phrase `\textit{plays guitar}', that is, the formula $!(N\bs S)$. One of the copies is then moved to the ellipsis site ``\textit{does too}", and  identified using the $(\bs L)$-rule.
	
	\subsection{Anaphora with Ellipsis}
		We now combine  anaphora and ellipsis and work with the example ``\textit{John likes his code, Bill does too.}", where we have an anaphor   `\textit{his}' and ellipsis marker `\textit{does too}'. Clearly, the derivation of the sloppy reading  will involve more resolutions than the strict reading, so one would expect this derivation to be more complex than a strict one, which is indeed the case.
			
			Below, we present the derivations of the strict and sloppy readings in full, using the following type-dictionary for both examples:
			\[
			\begin{array}{l}
			\{
				(\text{John}: !N), 
				(\text{likes}: !((!(!N\bs S))/N),
				(\text{his}: (!(!N\bs N))/N),
				(\text{code}: N),
				\\
				\quad (\text{Bill}: !N),
				(\text{does too}: (!(!N\bs S))\bs(!N\bs S))
			\}
			\end{array}\]
		\subsubsection*{Strict Reading.}
			
		The derivation of the strict reading is presented in figure \ref{fig:derivstrict}, where we can see only two uses of the $(contr)$-rule. We expect only two uses of  $(contr)$ since we are first resolving the anaphor `\textit{his}' with `\textit{John}', and then the verb-phrase `\textit{likes John's code}' with the ellipsis site `\textit{does-too}'.
		
			\begin{figure}[t]
			\[\scalebox{0.7}{
			\infer[(! L)]{!N, !((!(!N\bs S))/N), (!(!N \bs N))/N, N, !N, (!(!N\bs S))\bs (!N\bs S) \to S,S}
				{\infer[(/L)]{!N, (!(!N\bs S))/N, (!(!N \bs N))/N, N, !N, (!(!N\bs S))\bs (!N\bs S) \to S,S}
					{\infer[]{N \to N}{}
					&
					\infer[(!L)]{!N, (!(!N\bs S))/N, !(!N \bs N), !N, (!(!N\bs S))\bs (!N\bs S) \to S,S}
						{\infer[(contr)]{!N, (!(!N\bs S))/N, !N \bs N, !N, (!(!N\bs S))\bs (!N\bs S) \to S,S}
							{\infer[(perm_2)]{!N, !N, (!(!N\bs S))/N, !N \bs N, !N, (!(!N\bs S))\bs (!N\bs S) \to S,S}
								{\infer[(\bs L)]{!N, (!(!N\bs S))/N, !N, !N \bs N, !N, (!(!N\bs S))\bs (!N\bs S) \to S,S}
									{\infer[]{!N \to !N}{}
									&
									\infer[(/L)]{!N, (!(!N\bs S))/N, N, !N, (!(!N\bs S))\bs (!N\bs S) \to S,S}
										{\infer[]{N \to N}{}
										&
										\infer[(contr)]{!N, !(!N\bs S), !N, (!(!N\bs S))\bs (!N\bs S) \to S,S}
											{\infer[(perm_2)]{!N, !(!N\bs S), !(!N\bs S), !N, (!(!N\bs S))\bs (!N\bs S) \to S,S}
												{\infer[(!L)]{!N, !(!N\bs S), !N, !(!N\bs S), (!(!N\bs S))\bs (!N\bs S) \to S,S}
													{\infer[(\bs L)]{!N, !N\bs S, !N, !(!N\bs S), (!(!N\bs S))\bs (!N\bs S) \to S,S}
														{\infer[]{!N \to !N}{}
														&
														\infer[(\bs L)]{S, !N, !(!N\bs S), (!(!N\bs S))\bs (!N\bs S) \to S,S}
															{\infer[]{!(!N\bs S) \to  !(!N\bs S)}{}
															&
															\infer[(\bs L)]{S, !N, !N\bs S \to S,S}
																{\infer[]{!N \to !N}{} 
																& 
																\infer[]{S,S \to S,S}{}}}}}}}}}}}}}}
				}\]
				\caption{The derivation of the strict reading: `John likes John's code, Bill likes John's code', in $\blstar$}
				\label{fig:derivstrict}
				\end{figure}

	\subsubsection*{Sloppy Reading.}
	The derivation of the sloppy reading is presented in figure \ref{fig:derivsloppy}, and is clearly more complex than the strict reading in figure \ref{fig:derivstrict}. Looking at the right hand side of the sloppy derivation we see four instances of the $(contr)$-rule, which correspond to the four different coreferences we are resolving. The Gentzen presentation really obfuscates the coreference resolution visually, and one is forced to read these derivations very closely to understand what is happening. It is far easier to read the string-diagrams attached to these proofs which we demonstrate later in section \ref{sec:VSScoreference}
		\begin{figure}[H]
			\[\scalebox{0.7}{
			\infer[(/L)]{!N, !((!(!N\bs S))/N), (!(!N \bs N))/N, N, !N, (!(!N\bs S))\bs (!N\bs S) \to S,S}
				{\infer[]{N \to N }{}
				&
				\infer[(contr)]{!N, !((!(!N\bs S))/N), !(!N \bs N), !N, (!(!N\bs S))\bs (!N\bs S) \to S,S}
					{\infer[(contr)]{!N, !((!(!N\bs S))/N), !(!N \bs N), !(!N \bs N), !N, (!(!N\bs S))\bs (!N\bs S) \to S,S}
						{\infer[(contr)]{!N, !N, !((!(!N\bs S))/N), !(!N \bs N), !(!N \bs N), !N, (!(!N\bs S))\bs (!N\bs S) \to S,S}
							{\infer[(perm_2)]{!N, !N, !((!(!N\bs S))/N), !(!N \bs N), !(!N \bs N), !N, !N, (!(!N\bs S))\bs (!N\bs S) \to S,S}
								{\infer[(!L)]{!N, !((!(!N\bs S))/N), !N, !(!N \bs N), !(!N \bs N), !N, !N, (!(!N\bs S))\bs (!N\bs S) \to S,S}
									{\infer[(perm_2)]{!N, !((!(!N\bs S))/N), !N, !N \bs N, !(!N \bs N), !N, !N, (!(!N\bs S))\bs (!N\bs S) \to S,S}
										{\infer[(!L)]{!N, !((!(!N\bs S))/N), !N, !N \bs N, !N, !(!N \bs N), !N, (!(!N\bs S))\bs (!N\bs S) \to S,S}
											{\infer[(\bs L)]{!N, !((!(!N\bs S))/N), !N, !N \bs N, !N, !N \bs N, !N, (!(!N\bs S))\bs (!N\bs S) \to S,S}
												{\infer[]{!N \to !N}{}
												&
												\infer[(\bs L)]{!N, !((!(!N\bs S))/N), N, !N, !N \bs N, !N, (!(!N\bs S))\bs (!N\bs S) \to S,S}
													{\infer[]{!N \to !N}{}
													&
													\infer[(contr)]{!N, !((!(!N\bs S))/N), N, N, !N, (!(!N\bs S))\bs (!N\bs S) \to S,S}
														{\infer[(perm_2)]{!N, !((!(!N\bs S))/N),!((!(!N\bs S))/N), N, N, !N, (!(!N\bs S))\bs (!N\bs S) \to S,S}
															{\infer[(!L)]{!N, !((!(!N\bs S))/N), N, !((!(!N\bs S))/N), N, !N, (!(!N\bs S))\bs (!N\bs S) \to S,S}
																{\infer[(!L)]{!N, (!(!N\bs S))/N, N, !((!(!N\bs S))/N), N, !N, (!(!N\bs S))\bs (!N\bs S) \to S,S}
																	{\infer[(/L)]{!N, (!(!N\bs S))/N, N, (!(!N\bs S))/N, N, !N, (!(!N\bs S))\bs (!N\bs S) \to S,S}
																		{\infer[]{N \to N}{}
																		&
																		\infer[(/L)]{!N, !(!N\bs S), (!(!N\bs S))/N, N, !N, (!(!N\bs S))\bs (!N\bs S) \to S,S}
																			{\infer[]{N \to N}{}
																			&
																			\infer[(! L)]{!N, !(!N\bs S), !(!N\bs S), !N, (!(!N\bs S))\bs (!N\bs S) \to S,S}
																				{\infer[(\bs L)] {!N, !N\bs S, !(!N\bs S), !N, (!(!N\bs S))\bs (!N\bs S) \to S,S}
																					{\infer[]{!N \to !N}{}
																					&
																					\infer[(perm_2)]{S, !(!N\bs S), !N, (!(!N\bs S))\bs (!N\bs S) \to S,S}
																						{\infer[(\bs L)]{S, !N, !(!N\bs S), (!(!N\bs S))\bs (!N\bs S) \to S,S}
																							{\infer[]{!(!N\bs S) \to !(!N\bs S)}{}
																							&
																							\infer[(\bs L)]{S, !N, !N\bs S \to S,S}
																								{\infer[]{!N \to !N }{}
																								& 
																								\infer[]{S,S \to S,S}{}}}}}}}}}}}}}}}}}}}}}}
			}\]
		\caption{The derivation of the sloppy reading: `John likes John's code, Bill likes Bill's code, in $\blstar$}
		\label{fig:derivsloppy}
		\end{figure}

	Despite the difficulty of gaining quick qualitative information from the derivations, we have already shown that we can derive anaphora, ellipsis and anaphora with ellipsis in both of its possible strict and sloppy readings. The ability of our calculus to distinguish between the strict and sloppy readings is a desirable property, and more importantly, this distinction carries over into the vector space semantics, as we demonstrate next.
	
\section{Vector Space Semantics of $\blstar$}\label{sec:VSSForBlstar}

	In this section we introduce the vector space semantics of $\blstar$, as defined in our previous work \citep{mcpheat2020categorical}. We briefly summarise the categorical semantics in \ref{sec:catSem}, but  refer the reader to \citep{mcpheat2020categorical} for full technical details.  We will point out any practical use of the categorical semantics in the examples of this paper, but please note that  it is also not necessary to do so; the vector space semantics provides what we need for our linguistic motivations. We also introduce the diagrammatic calculus of  \citep{mcpheat2020categorical}  with the exception that here we omit the use of thick strings as to minimise the amount of string diagrammatic machinery present in this paper, as the diagrams are not the main contribution. This will be  in \ref{subsec:stringDiagrams}; the diagrammatic calculus is as a didactic tool to let the reader visualise the derivations in section \ref{sec:coreferenceInBlstar}. The semantics of these derivations is presented in \ref{sec:VSScoreference}. Before we define the semantics, we first need to understand tensor algebras, and comultiplications on them as these structures will allow us to define semantics for $!$ and the $(contr)$-rule.

	\subsection{Tensor Algebras}\label{sec:tensorAlgebras}
	We recall the definition of a tensor algebra, and show how to construct a Fermionic Fock space from it. Dualised versions of these constructions were used to interpret $!$ of full linear logic \citep{Blute1994}, and then for $\blstar$ \citep{mcpheat2020categorical}. We briefly  introduce tensor algebras, Fermionic Fock spaces and their duals below, and refer to \citep{mcpheat2020categorical} for further details on these constructions.
		
		\begin{definition}
			Given a vector space $V$, the \textbf{tensor algebra} $TV$ is defined as 
			\[TV := \bigoplus_{n\geq 0}V^{\otimes n} = \R \oplus V \oplus (V\otimes V) \oplus (V\otimes V\otimes V) \oplus \cdots.\]
			We call the terms $V^{\otimes n}$ the $n$-\textbf{th layer} of $TV$.
			
			There is a monoid structure on $TV$ given by a multiplication $m : TV \otimes TV \to TV$ defined by layer-wise concatenation i.e.
			\[m((v_1\otimes \cdots\otimes v_n)\otimes (w_1 \otimes \cdots \otimes w_m)) := v_1\otimes \cdots\otimes v_n\otimes w_1 \otimes \cdots \otimes w_m,\]
			and a unit $u: \R \longrightarrow TV$ given by $u(1) := 1$.
		\end{definition}

			It is in fact the case that $T$ is a free functor $\Vect_\R\longrightarrow\mathbf{Alg}_\R$. Thus defining a monad on $\Vect_\R$ by composing with the forgetful functor $\mathbf{Alg}_\R \longrightarrow \Vect_\R$, where $\mathbf{Alg}_\R$ is the category of real associative algebras.
		
		Clearly, $TV$ is infinite-dimensional for any nonzero vector space $V$ and an inappropriate choice for our semantics of $!$. Instead, we will interpret $!$ using Fermionic Fock Spaces, which have similar properties as Tensor algebras, but can be made finite dimensional. To define a Fermionic Fock Space we will need  alternating tensor products.
		\begin{definition}
			Given a (real) vector space $V$ with basis $(e_i)_{i\in I}$ respectively, we define the $n$-\textbf{fold alternating tensor product} $V^{\wedge n}$ of $V$ as:
				\[\overbrace{V \wedge V \wedge \cdots \wedge V}^{n\text{-times}} := V^{\otimes n}/U, \]
				where $U$ is the vector space spanned by vectors of the form
				\[(e_{i_1}\otimes e_{i_2}\otimes \cdots \otimes e_{i_n}) - \mathrm{sgn}(\sigma)(e_{\sigma(i_1)}\otimes e_{\sigma(i_2)}\otimes \cdots \otimes e_{\sigma(i_n)})\,. \]
				In the above,  $i_j \in I$ for each $1\leq j \leq n$ and each permuation $\sigma$ on $n$-symbols.
				Vectors in $V\wedge V$ are linear combinations of equivalence classes of simple tensors, denoted $e_{i_1}\wedge e_{i_2}\wedge \cdots \wedge e_{i_n}$. The key point in this definition is that the vector $e_{i_1}\wedge e_{i_2}\wedge \cdots \wedge e_{i_n}$ is equal to the vector $\mathrm{sgn}(\sigma)(e_{\sigma(i_1)}\wedge e_{\sigma(i_2)}\wedge \cdots \wedge e_{\sigma(i_n)})$. For instance, if $n=2$, we have that $e_1 \wedge e_2 = - e_2\wedge e_1$, since the sign of the permutation $(12)$ is $-1$.
		\end{definition}
		\subsubsection*{Note}\label{note:alternatingZero}
			Basis vectors in $V^{\wedge n}$ with repeated factors are zero. Consider a vector $e_{i_1}\wedge e_{i_2}\wedge \cdots \wedge e_{i_n}$ where WLOG the first two factors are the same: $e_{i_1}=e_{i_2}$. Thus we have the equality 
			\[\begin{array}{cl}
				e_{i_1}\wedge e_{i_2}\wedge \cdots \wedge e_{i_n}&= \mathrm{sgn}(12)(e_{(12)i_1}\wedge e_{(12)i_2}\wedge \cdots \wedge e_{(12)i_n})\\
				&= -e_{i_2}\wedge e_{i_1}\wedge \cdots \wedge e_{i_n}\\
				&= -e_{i_1}\wedge e_{i_2}\wedge \cdots \wedge e_{i_n}.
			\end{array}\]
			However if a vector equals its negative, it must have been zero to begin with, thus confirming that  $e_{i_1}\wedge e_{i_2}\wedge \cdots \wedge e_{i_n} = 0$.

		With the definition of alternating tensor products under our belts, we can quickly define the Fermionic Fock space construction in close analogy to the tensor algebra definition.
		\begin{definition}\label{def:FermFockSpace}
			Given a vector space $V$ its \textbf{Fermionic Fock space} is the vector space
			\[\F V := \bigoplus_{n\geq 0}V^{\wedge n} = \R \oplus V \oplus (V\wedge V) \oplus (V\wedge V\wedge V) \oplus \cdots.\]
		\end{definition}
		The space $\F V$ also has a monoidal structure, defined exactly as for $TV$, the only difference being that the multiplication on $\F V$ is alternating. $\F V$ is also known as the Grassmanian algebra of $V$, or the antisymmetric tensor algebra of $V$.

			$\F$ is also a free functor, this time of the form $\Vect_\R \longrightarrow \mathbf{AAlg}_\R$, where $\mathbf{AAlg}_\R$ is the category of alternating real associative algebras.

			We recall the well-known fact that given a finite dimensional vector space $V$, we can easily see that $\F V$ is also finite dimensional by applying the pigeonhole principle to the dimension of $V$ and the number of layers in $\F V$. Concretely, note that for any $n > \dim V$, we must repeat some factor of $e_{i_1}\wedge e_{i_2}\wedge \cdots \wedge e_{i_n} \in V^{\wedge n}$, and so by the note before definition \ref{def:FermFockSpace} we have $e_{i_1}\wedge e_{i_2}\wedge \cdots \wedge e_{i_n} = 0$. Since $V^{\wedge n}$ is spanned by such vectors, we conclude that $V^{\wedge n} = 0$. Thus, $\F V = \bigoplus_{n=0}^{\dim V} V^{\wedge n}$ for finite dimensional $V$.

			This finite dimensionality gives us an even nicer way to write down a comonoid structure on $\F(V)$, since for finite dimensional vector spaces $V$ with a chosen basis $(e_i)_{i\in I}$, we have $V \cong V^*$ by taking the set $(f_i)_{i\in I}$ as a basis for $V^*$, where every functional $f_j$ is defined on the basis of $V$ as $f_j(e_i) = \delta_{ij}$. Thus, $(\F V)^* \cong \F V$, meaning we can define the comultiplication on $\F V$.		
			Of course this formally carries little significance, but practically it is easier to work with elements of $\F V$ rather than duals. The dualising process  has a more fundamental importance to the categorical semantics of \citep{mcpheat2020categorical}, as outlined in \ref{sec:catSem}. 
			
			The assumption that we can choose a basis is informed by our application domain.  In other words, the  basis are given to us  for any application of this theory,. For instance, for obtaining (non-neural) word vectors, one often works with vector spaces built from lemmatised versions of term-term matrices after dropping  stop words \citep{GrefenSadrEMNLP,KartSadrCoNLL} or learn the vectors and tensors by neural or non-neural machine learning algorithms on a fixed set of features \citep{wijnholds-etal-2020,LMT}. This makes the availability of bases immediate.
			
			In fact, since we work exclusively with the category of finite dimensional vector spaces with fixed bases, one could equivalently proceed entirely in the category whose objects are finite sets (the bases) and whose morphisms are matrices. Here, we recover the vector space semantics as defined above via the free functor $F: \mathbf{finSet}\to \fdVect_\R$, and we can interpret $!$ using powersets. For an $n$-dimensional vector space $V$ with basis $X$ we have that $\dim(\F(V)) = 2^n = |\mathcal{P}(X)|$, so clearly the vector spaces $\F(V)$ and $F(\mathcal{P}(V))$ are isomorphic. However, since we make such explicit use of the monoidal structure on $\F(V)$, as shown below, the use of this isomorphism remains unclear. In particular, we make use of the the grading of $\F(V)$, and there is no canonical isomorphism between the grading of $\F(V)$ and a grading of $F(\mathcal{P}(V))$.

			The monoidal product $m: \F V\otimes \F V \to \F V$ is given by layerwise concatenation as it was for tensor algebras. Dualising this product gives us a coproduct $\Delta : \F V \to \F V\otimes \F V$, defined by mapping vectors $\tilde{v} \in \F V$ to all possible vectors in $\F V \otimes \F V$ that multiply to give $\tilde{v}$. Explicitly, this is:
			\[
			\Delta (\tilde{v}) := \sum_{\tilde{u},\tilde{w}\in \F V \atop \tilde{v} = m(\tilde{u}\otimes \tilde{w})} \tilde{u} \otimes \tilde{w}
			\]
			
			This mapping is hard to use in practice, since if $\dim V = n$ then, $\dim \F V = 2^n$,  making the comultiplication $\Delta$ a $2^n \times (2^n)^2$ matrix. Any practical applications of this would use $n\geq 100$ at the very least, making the mapping quite difficult to use. We overcome this problem by considering other comultiplications which we will define in \ref{sec:VSScoreference}.

	\subsection{String Diagrams}\label{subsec:stringDiagrams}
	The formalisation of categorical reasoning using string diagrams was first achieved in \citep{JOYAL199155},  then systematically extended in \citep{Selinger2010} to autonomous, braided and traced categories, to name only a few. In the same style, \citep{BaezStay2011} interpret monoidal biclosed categories in their clasped diagrammatic calculus.
	 In this section, we go over  the clasp diagrams in order to later represent our derivations in a more legible format.   The use of clasp diagrams to depict Lambek calculus derivations is possible, since  the categorical semantics of Lambek calculus is a monoidal bi-closed category. This was initiated in  the work of \citep{Coeckeetal2013} and proved coherent in \citep{Wijnholds2017WoLLiC}.  In \citep{mcpheat2020categorical}, we added some more structure in the form of $\Delta$-maps and the diagrammatic $!$-modality. This extension of the clasp diagrams requires a new proof of coherence, which constitutes work in progress.

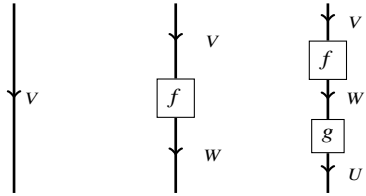
\begin{figure}
				\[
				{%
\beginpgfgraphicnamed{obsHomsComp}
\InputIfFileExists{obsHomsComp.tikz}{}{\input{./tikz/obsHomsComp.tikz}}
\endpgfgraphicnamed}
				\] 
				\caption{Vector spaces, linear maps and composition of linear maps, in terms of string diagrams}\label{fig:catDiagrams}
			\end{figure}

	We recall how to draw objects morphisms and composition diagrammatically in figure \ref{fig:catDiagrams}.
			
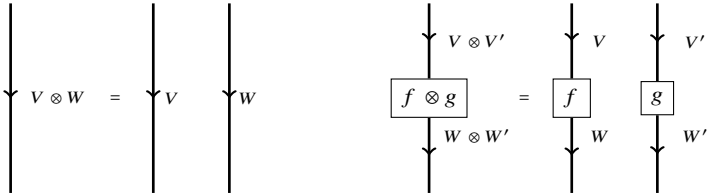
\begin{figure}
				\[
				{%
\beginpgfgraphicnamed{monObsHoms}
\InputIfFileExists{monObsHoms.tikz}{}{\input{./tikz/monObsHoms.tikz}}
\endpgfgraphicnamed}
				\] 
				\caption{Tensor products of vector spaces on the left, and of linear maps on the right.}\label{fig:tensorProducts}
			\end{figure}

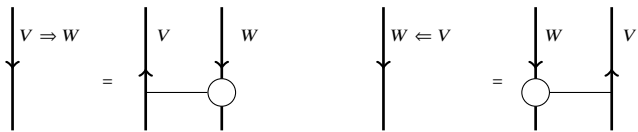
\begin{figure}
				\[
				{%
\beginpgfgraphicnamed{clasps}
\InputIfFileExists{clasps.tikz}{}{\input{./tikz/clasps.tikz}}
\endpgfgraphicnamed}
				\] 
				\caption{Right facing and left facing clasp diagrams representing vector spaces containing $\Rightarrow,\Leftarrow$.}\label{fig:clasps}
			\end{figure}

			Tensor products of vector spaces and linear maps are drawn side by side as in figure \ref{fig:tensorProducts}. We use the clasp notation of \citep{BaezStay2011} to draw vector spaces of the form $V \Rightarrow W$ and $W\Leftarrow V$ as depicted in figure \ref{fig:clasps}. Recall that these are the set of linear maps $V\to W$, which gets its vector space structure pointwise.		

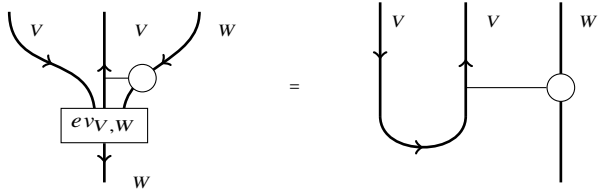
\begin{figure}
				\[
				{%
\beginpgfgraphicnamed{cups}
\InputIfFileExists{cups.tikz}{}{\input{./tikz/cups.tikz}}
\endpgfgraphicnamed}
				\]
				\caption{(Left) evaluation drawn as a cup}
				\label{fig:cups}
			\end{figure}

			Recall that evaluation of linear maps is a linear map itself of the form $\ev_{V,W}: V \otimes (V\Rightarrow W) \to W :: v\otimes f \mapsto f(v)$. We have a concise depiction of evaluation of linear maps in our diagrams, namely as ``cups", as seen in figure 
			\ref{fig:cups}. Note also that $\Leftarrow$ and $\Rightarrow$ are isomorphic, and so there is only one evaluation map. This is easiest to see by considering the two (a priori distinct) evaluation maps $v\otimes f \mapsto f(v)$ and $f\otimes v\mapsto f(v)$. Since the tensor product is symmetric, we can get the first map from the other by first applying symmetry and vice versa, making the two maps isomorphic.
			We choose to distinguish between $\Rightarrow$ and $\Leftarrow$ in our presentation of the semantics because it keeps us closer to the $\blstar$-syntax. Work is being done to define a tensor product on vector spaces which is non-symmetric, thus distinguishing between $\Rightarrow$ and $\Leftarrow$ in \citep{correia2020density}.
			
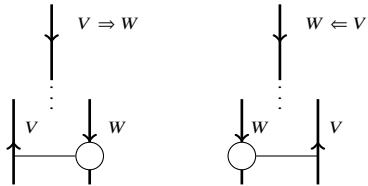
\begin{figure}\[
			{%
\beginpgfgraphicnamed{vdots}
\InputIfFileExists{vdots.tikz}{}{\input{./tikz/vdots.tikz}}
\endpgfgraphicnamed}
			\]
			\caption{How we identify strings with complex labels and their diagrams}\label{fig:vdots}
			\end{figure}

			In some cases it will simplify our diagrams significantly if we can interchange between our string diagrammatic conventions for  vector spaces defined using the $\Rightarrow$ and $\Leftarrow$ operations. We will denote these equalities using vertical dots as shown in figure \ref{fig:vdots}.
				
	\subsection{Vector Space Semantics for $\blstar$}\label{sec:semanticsDefn}
		We can now define the part of the vector space semantics for $\blstar$ necessary for the derivations of anaphora and ellipsis. We also show how to interpret the rules of $\blstar$ using the diagrams of the previous section,  which will enable us to read the derivations of anaphora and ellipsis in a far more legible manner. 
		For a definition of vector space semantics for the rest of $\blstar$ and the proof of soundness, we refer the reader to the original paper \citep{mcpheat2020categorical}.
		\begin{definition}
			We inductively define a semantic map, which using conventional notation we denote by  `` $\semantics{ \ }$", on formulas and derivable sequents of $\blstar$. This map sends  formulas to  finite dimensional real vector spaces, and derivable sequents to linear maps. We give the inductive definition below. 
			
			\begin{itemize}
			\item To atomic types $N$ and $S$ of $\blstar$ we assign vector spaces as follows 
			\[
			\semantics{N}:= V_N \qquad \semantics{S}:= V_S
			\]
			\item To complex types of $\blstar$ we assign: 
			\[\begin{array}{ccl}
				\semantics{A,B} &:=& \semantics{A} \otimes \semantics{B} \\
				\semantics{A \bs B} &:=& \semantics{A} \Rightarrow \semantics{B} \\
				\semantics{B / A} &:=& \semantics{B} \Leftarrow \semantics{A} \\
				\semantics{!A} &:=& \F\semantics{A}
			\end{array}\]
			\end{itemize}
				\end{definition}
			In the above definition,  for any two vector spaces $V,W$, the spaces $V\Rightarrow W$ and $W \Leftarrow V$ denote the set of linear maps from $W$ to $V$. The space $\F V$ for a finite dimensional vector space $V$ is the Fermionic Fock Space of $V$.
			For a finite list of formulas $\Gamma =\{A_1,A_2,\ldots,A_n\}$ we define $\semantics{\Gamma}:= \semantics{A_1} \otimes \semantics{A_2} \otimes \cdots\otimes \semantics{A_n}$.

			Derivable sequents $\Gamma \longrightarrow A$ are interpreted as linear maps $\semantics{\Gamma} \longrightarrow \semantics{A}$. Since sequents are not typically labelled, we add lower case roman letters ($f,g,h,\ldots$) to name linear maps when needed.
			To define the interpretation a derivable sequent $\Gamma \longrightarrow A$ in practice, one essentially builds it from the root of the derivation up, following the below interpretations of the proof rules of $\blstar$. As we introduce the semantics of all the rules of $\blstar$ we also show how to depict them as string diagrams in $\fdVect_\R$.
			
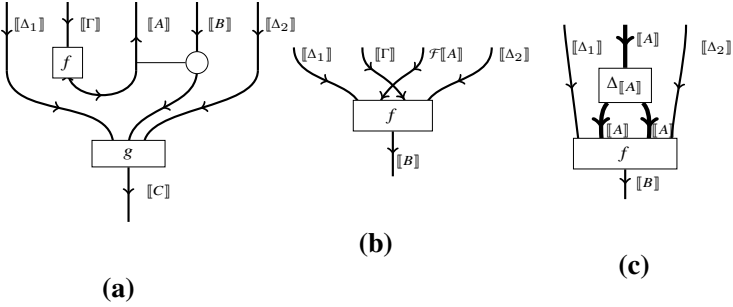
\begin{figure}[H]
				\centering
				\begin{subfigure}{0.28\textwidth}
					\[\scalebox{0.8}{{%
\beginpgfgraphicnamed{backslashL}
\InputIfFileExists{backslashL.tikz}{}{\input{./tikz/backslashL.tikz}}
\endpgfgraphicnamed}}\quad\]
					\caption{}\label{fig:backslashL}
				\end{subfigure}
				\begin{subfigure}{0.28\textwidth}
					\[\quad\scalebox{0.8}{{%
\beginpgfgraphicnamed{perm2}
\InputIfFileExists{perm2.tikz}{}{\input{./tikz/perm2.tikz}}
\endpgfgraphicnamed}}\quad\]
					\caption{}\label{fig:perm}
				\end{subfigure}
				\begin{subfigure}{0.28\textwidth}
					\[\quad\scalebox{0.8}{{%
\beginpgfgraphicnamed{contr2}
\InputIfFileExists{contr2.tikz}{}{\input{./tikz/contr2.tikz}}
\endpgfgraphicnamed}}\]
					\caption{}\label{fig:contr}
				\end{subfigure}
				\caption{Diagrammatic interpretation of structural $\blstar$ rules.}
			\end{figure}

			We begin by interpreting the axiom rule $A\longrightarrow A$ of $\blstar$. This is simply interpreted as the existence of an identity map $id_{\semantics{A}}: \semantics{A} \longrightarrow \semantics{A}$. Diagrammatically, the axiom rule is depicted with a string labelled $\semantics{A}$.
			%
			The $(\bs L)$ and $(/L)$-rules are interpreted very similarly, so we will show one and let the reader deduce the other. We recall the syntax of the $(\bs L)$-rule on the left below, and lay out the semantics on the right and define it below
			\[\scalebox{0.9}{
			\infer[(\bs L)]
				{\Delta_1, \Gamma, A\bs B, \Delta_2 \longrightarrow C}
				{\Gamma \longrightarrow A
				&
				\Delta_1, B, \Delta_2 \longrightarrow C}
			\quad
			\infer[(\bs L)]
				{h:\semantics{\Delta_1}\otimes \semantics{\Gamma}\otimes \semantics{A}\Rightarrow \semantics{B}\otimes \semantics{\Delta_2} \longrightarrow \semantics{C}}
				{f: \semantics{\Gamma} \longrightarrow \semantics{A}
				&
				g: \semantics{\Delta_1}\otimes \semantics{B}\otimes \semantics{\Delta_2} \longrightarrow \semantics{C}}	
			}\]
			where we are given $f$ and $g$, and define $h$ as 
			\[h := g \circ (id_{\Delta_1} \otimes \ev_{\semantics{A},\semantics{B}} \otimes id_{\Delta_2} )\circ (id_{\Delta_1} \otimes f \otimes id_{\semantics{A}\Rightarrow \semantics{B}} \otimes id_{\Delta_1})
			\]
			which can be visualised diagrammatically in figure \ref{fig:backslashL}.
			
		Next we have the semantics of the permutation rules, which are immediate from the symmetry of the tensor product in $\fdVect_\R$.
		Diagrammatically, these rules say that we may cross strings representing Fock spaces as done in figure \ref{fig:perm}.
		
		Finally, we have the contraction rule. This is interpreted using the comonoidal comultiplication introduced in \ref{sec:tensorAlgebras}.
		\[
		\infer[(contr)]
			{\Delta_1, !A,\Delta_2 \longrightarrow B }
			{\Delta_1, !A, !A, \Delta_2 \longrightarrow B}
		\quad
		\infer[(contr)]
			{f^c: \semantics{\Delta_1} \otimes \semantics{!A}  \otimes \semantics{\Delta_2} \longrightarrow \semantics{B} }
			{f: \semantics{\Delta_1} \otimes \semantics{!A} \otimes \semantics{!A} \otimes \semantics{\Delta_2} \longrightarrow \semantics{B} }	
		\]
		where given $f$, we define $f^c$ as 
		\(
		f^c := f \circ (id_{\semantics{\Delta_1}} \otimes \Delta_{\semantics{A}} \otimes id_{\semantics{\Delta_2}}),
		\)
		where $\Delta_{\semantics{A}}$ is a comultiplication $\F \semantics{A} \to \F \semantics{A} \otimes \F \semantics{A}$. Diagrammatically, this rule lets us `split' strings corresponding to $!$-ed formulas, as in figure \ref{fig:contr}.
		
		Note that the $\Delta$-box only specifies the type of the map $\Delta$, meaning that we can do a derivation in $\blstar$, interpret it as string diagram, and then choose whichever instances of $\Delta$ as we please.
		
		It is worth pointing out that we do not aim to define a  \textit{complete} model. As probably already noted by the reader,  interpreting $\blstar$ in terms of vector spaces could never be complete; take for example the distinction between $\bs $ and $/$ in $\blstar$ which is clearly not carried into the semantics. However, this is common practice in the DisCoCat line of research \citep{Coeckeetal2010,Coeckeetal2013}  and has also been employed in set theoretic semantics of Lambek calculus \citep{VanBenthem1988}. 
		The choice of vector spaces as an interpretion of $\blstar$ reflects that one can convey the same meanings in syntacically different languages.  Consider for example the English sentence "\textit{John likes Mary}" and the Farsi sentence "\textit{John Mary-ra Doost-darad(likes)}". The syntaxes are distinct, but the meanings are the same. This is despite the fact that the order of words is very different in the two languages. 
		
		Finding complete models of $\blstar$ is an interesting pursuit, although not the goal of the current paper. Work in this direction has begun in \citep{correia2020density,Grecoetal2020}, where the use of different nonsymmetric tensor products are investigated. One of our  reviewers has suggested specifying subspaces of  atomic vector spaces in order to distinguish between spaces and their duals. This also takes us closer to a complete model. We have also considered defining a model in $C^*$-algebras, which have a nonsymmetric tensor product.

	\subsection{Categorical Vector Space Semantics}\label{sec:catSem}
		For readers who are familiar with categorical semantics of linear logic such as \citep{Mellies2014}, and diagrammatic reasoning, you may have noticed some category-theoretic constructions, which we have not mentioned; we briefly outline them here and again refer the reader to \citep{mcpheat2020categorical} for full detail. The category theoretic machinery is incredibly useful and succinct for proving soundness results, but can be cumbersome in practice when looking at concrete applications as we do in this paper.
		
		We mention that $\F$ is a free functor $\fdVect_\R \to \mathbf{AAlg}_\R$, and then mention dualising spaces $\F V$. Combining these two facts actually defines comonad on $\fdVect_\R$. First of all, since $\F$ is free, we have a a right adjoint forgetful functor $\mathbf{AAlg}_\R \to \fdVect_\R$, which when composed form a \textit{monad} $U\F :\fdVect_\R \to \fdVect_\R$. Then, pre and post-compose the functor $U\F$ with the vector space dual, which as shown in \citep{HopfMonads}, defines a comonad on $\fdVect_\R$, not to be confused with the comonad on $\mathbf{AAlg}_\R$, induced by the adjunction.
		
		To explicitly define the structure of this comonad we will make use of the isomorphism $(U\F V^*)^* \cong U\F V$ to simplify the mathematics.
		The counit of the comonad $(\epsilon_V : U\F V \to V)_{V\in \fdVect_\R}$ is defined using projection onto the first layer. The comultiplication of the comonad $(\delta_{V}:U\F V \to U\F U \F V)_{V\in \fdVect_\R}$ is given by inclusion into the first layer. That is, $\epsilon_V(v_0,v_1,v_2\wedge v_3, \ldots) = v_1 \in V$, and $\delta_V(\tilde{v}) := (0,\tilde{v},0,0,\ldots)$

		There is an alternative categorical interpretation of the relevant modality $!$ provided by \citep{Jacobs1994} using what the author calls ``relevant monads''. However, these monads are the categorical semantics of a smaller fragment of logic which is only concerned with contraction, as opposed to $\blstar$ which considers $!$ to be responsible for both contraction and permutation. Further, for the semantics of  $!$ to be sound for the full logic of $\blstar$, we need  $!$ to be a comonad rather than a monad. The comonad counit propoerty is necessary to prove soundness of the $(!L)$-rule of the calculus.  This is the approach taken by the paper  \citep{Blute1994} showing that Fock spaces can been used to interpret the linear logic $!$-modality and provide a sound semantics for the full logic. There is a natural connection to a comonadic interpretation rather than a monadic one, again suggesting that relevant monads are not the immediate best semantics for the $!$ of $\blstar$. However considering a smaller fragment of $\blstar$, or perhaps a Lambek calculus with a soft exponential modality as in \citep{Lafont2004} may allow the use of contraction monads. 
		
\section{Examples of Vector Space Semantics of Anaphora and Ellipsis}\label{sec:VSScoreference}
	In this section we show how to interpret each of the derivations from section \ref{sec:coreferenceInBlstar} in our vector space semantics. In each example we first draw the string diagram corresponding to the derivation which firstly gives us a far more readable version of the sequent derivation, and secondly gives us  the linear map corresponding to the meaning. We demonstrate how to extract the linear map from the diagram, and show how to evaluate it for each example.
	
	We will use the following notation in the coming subsections. Superscript tildes denote vectors in Fock spaces (i.e. $\tilde{v}\in \F V$). 
	We fix bases $(n_i)_{i\in I}$ and $(s_j)_{j\in J}$ for spaces $\semantics{N}, \semantics{S}$ respectively. Basis elements  marked with asterisks denote basis elements in dual spaces $\semantics{N}^*, \semantics{S}^*$, i.e. $n_i^*: \semantics{N}\to \R :: n_{i'} \mapsto \delta_{}ii'$.
	We also recall that you may define a basis of a tensor product of vector spaces $V\otimes W$, as the tensor products of the basis vectors of $V$ and $W$. In particular for spaces of the form $\semantics{N}\Rightarrow \semantics{S}$ we work with the basis $(n_i^*\otimes s_j)_{i \in I, j\in J}$. We also fix a number $k\in\R$ and use boldface $\mathbf{k}$ to denote a vector of $k$'s in the appropriate vector space, e.g. $\mathbf{k} = (k,k)\in \R^2$ or $\mathbf{k}=(k,k,k,k)\in \R^4$ and so on.
	
	In the following examples we will write out the semantics of the derivations from section \ref{sec:coreferenceInBlstar} using two different instances of the $\Delta$-map. However there is a way to write out the semantics without specifying a $\Delta$-map, by using \textbf{Sweedler notation}. This is a notation for abstract comultiplication maps $\Delta: V\to V\otimes V$ where we write $\Delta(v) := v_{(1)}\otimes v_{(2)}$. Once $\Delta$ is specified, we substitute the relevant terms in for $v_{(1)}$ and $v_{(2)}$. 
	We use this notation when appropriate to simplify or generalise the computations. 
	The $\Delta$-maps we use are called the $\mathbf{k}$-\textbf{extension} and \textbf{basis copy} maps. $\mathbf{k}$-extension maps vectors $v$ to $v\otimes \mathbf{k} + \mathbf{k} \otimes v$ (this was called \textit{cofree-inspired} in \citep{mcpheat2020categorical}). The basis copy map is defined on the basis of the relevant vector space as $e_i \mapsto e_i\otimes e_i$, and extended linearly to the whole space. 
	After extending, one can give two mathematically equivalent versions of this map by gathering the coefficients to the left or right factor. That is, if we apply basis copying to a vector $v = \sum_iC_iv_i$ we get $\Delta(v) = \sum_iC_i (v_i\otimes v_i)$, and by gathering the coefficients on the left factor we get $\Delta(v) = v\otimes \mathbf{1}$, or on the right we get $\Delta(v) = \mathbf{1} \otimes v$. Although mathematically equivalent, a choice must be made to implement this model. We call the maps gathering the coefficients on the left(right) basis copying a(b) (these were called \textit{cogebra a(b)} in \citep{mcpheat2020categorical}).
	
	\subsection{Semantics of Anaphora}\label{subSec:anaphoraSemantics}
\begin{figure}
			\[\scalebox{0.9}{{%
\beginpgfgraphicnamed{JohnSleeps1}
\InputIfFileExists{JohnSleeps1.tikz}{}{\input{./tikz/JohnSleeps1.tikz}}
\endpgfgraphicnamed}}\]
			\caption{Anaphora diagram}\label{fig:anaphoraDiagram}
		\end{figure}
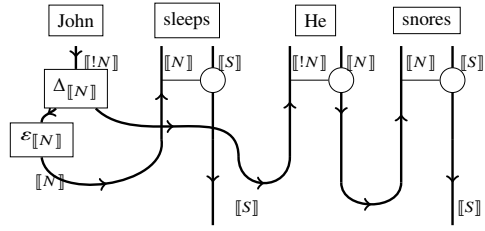

	We begin with drawing the diagram corresponding to the derivation in figure \ref{fig:anaphora}. Reading the derivation from bottom to top, we draw the diagram from top to bottom\footnote{For a step-by-step example of how to draw these diagrams, we refer the reader to \citep{mcpheat2020categorical} where we present how to draw a string diagram from a $\blstar$-derivation.}.
	The resulting string diagram in figure \ref{fig:anaphoraDiagram}, clearly shows that the meaning of `\textit{John}' is being copied by the $\Delta_{\semantics{N}}$ map, and then one of the copies is sent to `\textit{He}', which is in turn sends the meaning of `\textit{John}' to the input of `\textit{snores}'. This is far easier to see using string diagrams rather than the sequent calculus derivation.
	The vector spaces at the top and the bottom of the diagram (corresponding respectively to the semantics of the left and right hand sides of the sequent $!N,N\bs S, !N\bs N, N\bs S \longrightarrow S,S$) tell us that this diagram defines a linear map, say $f$, of type 
	\small\[
	\F \semantics{N} \otimes \semantics{N} \otimes \semantics{S}\otimes \F\semantics{N}\Rightarrow \semantics{N} \otimes \semantics{N}\Rightarrow\semantics{S} \to \semantics{S} \otimes \semantics{S}\,.
	\]\normalsize
	Explicitly, taking the vectors $\widetilde{John} \in \F\semantics{N}$, $\overrightarrow{sleeps},\overrightarrow{snores} \in\semantics{N}\Rightarrow\semantics{S}$ and $\overrightarrow{He} \in \F\semantics{N}\Rightarrow \semantics{N}$, we may define $f$ (in Sweedler notation) as:
	\[f(\widetilde{John} \otimes \overrightarrow{sleeps} \otimes \overrightarrow{He}\otimes \overrightarrow{snores}):= 
	\overrightarrow{sleeps}(\epsilon_{\semantics{N}}(\widetilde{John}_{(1)})) \otimes \overrightarrow{snores}(\overrightarrow{He}(\widetilde{John}_{(2)}) ).\]
	By using $\mathbf{k}$-extension or basis copying for the $\Delta_{\semantics{N}}$-map, $f$ becomes:
	\small\[
	\begin{array}{l}
\mathbf{k}\text{\bf-extension}:\\
	 f(\widetilde{John} \otimes \overrightarrow{sleeps} \otimes \overrightarrow{He}\otimes \overrightarrow{snores}):=\\
	 \qquad
	\overrightarrow{sleeps}(\epsilon_{\semantics{N}}(\widetilde{John}))\otimes \overrightarrow{snores}(\overrightarrow{He}(\mathbf{k})) 
		+
		\overrightarrow{sleeps}(\epsilon_{\semantics{N}}(\mathbf{k}))\otimes \overrightarrow{snores}(\overrightarrow{He}(\widetilde{John})) 
\\
\text{\bf Basis copy}: \\
	 f(\tilde{n}_i \otimes (n_{i'}^*\otimes s_j) \otimes (\tilde{n}_{i''}^* \otimes n_{i'''}) \otimes (n_{i''''}^* \otimes s_{j'}))
	:=
 n_{i'}^*(n_i)s_j \otimes \tilde{n}^*_{i''}(\tilde{n}_{i})n_{i''''}^*(n_{i'''})s_{j'}
	\end{array}
	\]\normalsize

	\subsection{Semantics of Ellipsis}\label{subSec:ellipsisSemantics}
	\label{subsec:vectellipsis}
	We proceed in this example as we did for the anaphora example, by reading the derivation of the ellipsis (figure \ref{fig:ellipsis}) from the root, and the diagram from the top, as presented in figure \ref{fig:ellipsisDiagram}.
		\begin{figure}[t!]
			\[\scalebox{0.9}{{%
\beginpgfgraphicnamed{ellipsis}
\InputIfFileExists{ellipsis.tikz}{}{\input{./tikz/ellipsis.tikz}}
\endpgfgraphicnamed}}\]
			\caption{Ellipsis string diagram}\label{fig:ellipsisDiagram}
		\end{figure}
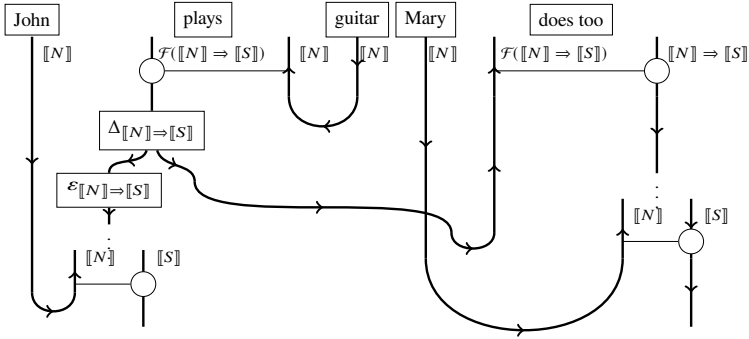
				\begin{figure}[t!]
			\[\scalebox{0.9}{{%
\beginpgfgraphicnamed{JohnLikesHisCode}
\InputIfFileExists{JohnLikesHisCode.tikz}{}{\input{./tikz/JohnLikesHisCode.tikz}}
\endpgfgraphicnamed}}\]
			\caption{Strict Diagram}\label{fig:strictDiagram}
		\end{figure}
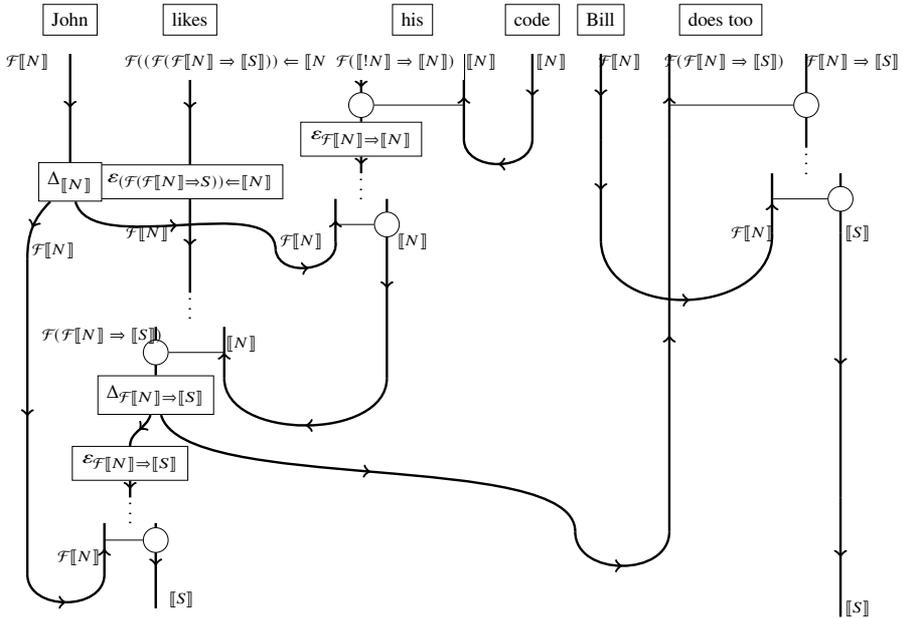
		By inspecting the strings at the top and bottom of the diagram, we see that this defines a linear map, say $g$, of type
	\[\scriptstyle
	\semantics{N} \otimes (\F(\semantics{N} \Rightarrow \semantics{S}))\Leftarrow \semantics{N}\otimes \semantics{N} \otimes \semantics{N} \otimes \F(\semantics{N} \Rightarrow \semantics{S})\Rightarrow (\semantics{N} \Rightarrow \semantics{S} ) \to \semantics{S} \otimes \semantics{S}. 
	\]
	If we consider vectors 
	\small\[\begin{array}{l}
	\overrightarrow{John},\overrightarrow{guitar},\overrightarrow{Mary} \in \semantics{N}, \quad
	\overrightarrow{plays}\in \F(\semantics{N}\Rightarrow \semantics{S})\Leftarrow \semantics{N},
	\\
	\qquad\qquad
	\overrightarrow{does\,too} \in \F(\semantics{N}\Rightarrow \semantics{S}) \Rightarrow (\semantics{N}\Rightarrow \semantics{S})
	\end{array}\]\normalsize
	we can define $g$ as
	\small\[
	\begin{array}{l}
	g(\overrightarrow{John}\otimes \overrightarrow{plays}\otimes \overrightarrow{guitar}\otimes \overrightarrow{Mary}\otimes \overrightarrow{does\,too}) 
	:=\\
	\quad
	\epsilon_{\semantics{N}\Rightarrow\semantics{S}}((\overrightarrow{plays}(\overrightarrow{guitar})_{(1)}) (\overrightarrow{John})\otimes\overrightarrow{does\,too}(\overrightarrow{plays}(\overrightarrow{guitar})_{(2)},\overrightarrow{Mary}).
	\end{array}
	\]\normalsize
	Note that  function $\overrightarrow{plays}$ is a function of type $\semantics{N}\to \F(\semantics{N}\Rightarrow \semantics{S})$ and is first being evaluated at the noun $\overrightarrow{guitar}$. This lets us define 
	\[
	\widetilde{plays\,guitar}\in \F(\semantics{N}\Rightarrow \semantics{S}) \quad  \mbox{as} \quad  \widetilde{plays \, guitar}:= \overrightarrow{plays}(\ov{guitar})\,.
	\]
	We may also consider taking $\overrightarrow{does\,too}$ to be a projection, like $\epsilon$, which maps $\widetilde{plays\, guitar}$ to a vector $\overrightarrow{plays\,guitar}\in \semantics{N}\Rightarrow \semantics{S}$. This makes the definition of $g$ slightly neater:
	\small\[
	\begin{array}{l}
	g(\overrightarrow{John}\otimes \overrightarrow{plays}\otimes \overrightarrow{guitar}\otimes \overrightarrow{Mary}\otimes \overrightarrow{does\,too})
	:= \\ \quad
	\epsilon_{\semantics{N}\Rightarrow\semantics{S}}(\widetilde{plays\,guitar}_{(1)})(\overrightarrow{John})
	\otimes
	\overrightarrow{plays\,guitar}(\overrightarrow{Mary}).
	\end{array}\]\normalsize
	The mathematical form of $g$ looks very similar to the natural language semantics `\textit{John plays guitar, Mary does too}'.

	Finally, when specifying the $\Delta$-maps to be one of the $\mathbf{k}$-extension and basis copying we get:
		\small\[
		\begin{array}{l}
		\mathbf{k}\text{\bf-extension}\\ \,g(\overrightarrow{John} \otimes \overrightarrow{plays} \otimes \overrightarrow{guitar}\otimes \overrightarrow{Mary} \otimes \overrightarrow{does\,too}) :=
		\\
		\qquad \epsilon_{\semantics{N}\Rightarrow\semantics{S}}((\overrightarrow{plays}(\overrightarrow{guitar})) (\overrightarrow{John})\otimes\overrightarrow{does\,too}(\mathbf{k},\overrightarrow{Mary}) +
		\\ 
		\qquad \quad \epsilon_{\semantics{N}\Rightarrow\semantics{S}}(\mathbf{k}) (\overrightarrow{John})\otimes\overrightarrow{does\,too}(\overrightarrow{plays}(\overrightarrow{guitar},\overrightarrow{Mary})
		\\
		\text{\bf Basis copy}\\ g(
		n_{i_1} \otimes (\widetilde{n_{i_2}^* \otimes s_{j_1}}\otimes n_{i_3}^*)\otimes n_{i_4} \otimes n_{i_5} \otimes ((\widetilde{n_{i_6}^*\otimes s_{j_2}})^*\otimes n_{i_7}^*\otimes s_{j_3})
		):=
		\\
		\qquad n_{i_2}^*(n_{i_1})s_j n_{i_3}^*(n_{i_4})
		\otimes 
		(\widetilde{n_{i_6}^*\otimes s_{j_2}})^*(\widetilde{n_{i_2}^* \otimes s_{j_1}})n_{i_7}^*(n_{i_5})s_{j_3}.
		\end{array}\]\normalsize

	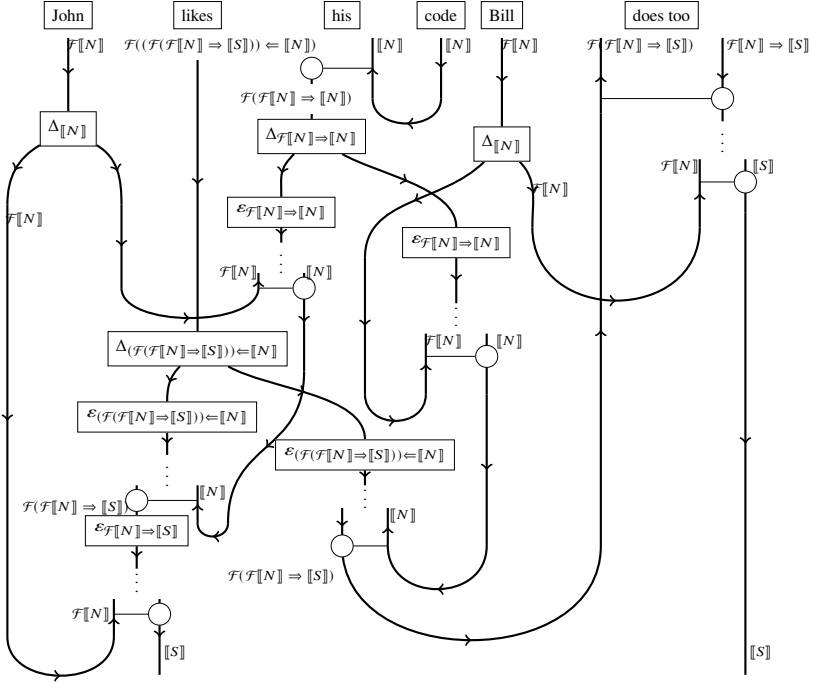
\begin{figure}[t!]
		 \[\scalebox{0.8}{{%
\beginpgfgraphicnamed{JohnLikesHisCodeSloppy}
\InputIfFileExists{JohnLikesHisCodeSloppy.tikz}{}{\input{./tikz/JohnLikesHisCodeSloppy.tikz}}
\endpgfgraphicnamed}}\]
		 \caption{Sloppy Diagram}\label{fig:sloppyDiagram}
	\end{figure}

	\subsection{Anaphora with Ellipsis}
	
	An important contribution to note in the following is that the linear maps for the strict and sloppy semantics are distinct, thus showing that not only does $\blstar$ provide a syntactic distinction between strict and sloppy, but also so does the vector space semantics. This overcomes the difficulty faced in \citep{WijnSadr2019} where there are indeed two different syntactic derivations, but the corresponding vector spaces semantics collapses and they become equal to each other. 

	As we have done in \ref{subSec:anaphoraSemantics} and \ref{subSec:ellipsisSemantics} we first present the diagrams of the strict and sloppy derivations of (figures \ref{fig:derivstrict} and \ref{fig:derivsloppy}), which is done in figures \ref{fig:strictDiagram} and \ref{fig:sloppyDiagram} respectively.

	Again, by inspecting the wires on the top and bottom of both diagrams, we see that they specify linear maps, say $h_{strict}, h_{sloppy}$, which are both of the following type:
	\[\begin{array}{l} \scriptstyle
	\F\semantics{N}\otimes 
	\F(\F(\F\semantics{N}\Rightarrow\semantics{S})\Leftarrow\semantics{N}) \otimes 
	\F(\F\semantics{N}\Rightarrow\semantics{N})\Leftarrow\semantics{N}\otimes 
	\semantics{N}\otimes \\  
		\scriptstyle\qquad
		\F\semantics{N}\otimes 
		(\F(\F\semantics{N}\Rightarrow \semantics{S}))\Rightarrow (\F\semantics{N}\Rightarrow \semantics{S}) 
		\to \semantics{S}\otimes \semantics{S} 
	.\end{array}\]
	
	Taking vectors
	\small\[\begin{array}{c}
	\widetilde{John},\widetilde{Bill}\in\F\semantics{N},
	\quad
	\widetilde{likes}\in \F(\F(\F\semantics{N}\Rightarrow\semantics{S})\Leftarrow\semantics{N})
	\\
	\overrightarrow{his}\in \F(\F\semantics{N}\Rightarrow\semantics{N})\Leftarrow\semantics{N}, \quad
	\overrightarrow{code}\in \semantics{N}
	\\
	\qquad 
	\overrightarrow{does\,too}\in (\F(\F\semantics{N}\Rightarrow \semantics{S}))\Rightarrow (\F\semantics{N}\Rightarrow \semantics{S})
	\end{array}\]\normalsize

	we can define the maps $h_{strict}, h_{sloppy}$ with Sweedler notation as:
	\small\[
	\begin{array}{l}
	h_{strict}(
		\widetilde{John} \otimes 
		\widetilde{likes} \otimes
		\overrightarrow{his} \otimes
		\overrightarrow{code} \otimes
		\widetilde{Bill} \otimes 
		\overrightarrow{does\,too}
	) :=
	\\ \
	\epsilon_{\F\semantics{N}\Rightarrow\semantics{S}}(\epsilon_{\F(\F\semantics{N}\Rightarrow\semantics{S})\Leftarrow\semantics{N}}(\widetilde{likes})(\epsilon_{\F\semantics{N}\Rightarrow\semantics{N}}(\widetilde{his}(code)))_{(1)})(\widetilde{John}_{(2)}) \otimes 
	\\
	\qquad
	\overrightarrow{does\,too}(
		\epsilon_{\F(\F\semantics{N}\Rightarrow\semantics{S})\Leftarrow\semantics{N}}(\widetilde{likes})(\epsilon_{\F\semantics{N}\Rightarrow\semantics{N}}(\widetilde{his}(code)))_{(2)}))(\epsilon_{\semantics{N}})(\widetilde{Bill}))
	\\ \\
	h_{sloppy}(
		\widetilde{John} \otimes 
		\widetilde{likes} \otimes
		\overrightarrow{his} \otimes
		\overrightarrow{code} \otimes
		\widetilde{Bill} \otimes 
		\overrightarrow{does\,too}
	) :=
	\\ \
	\epsilon_{\F(\F\semantics{N}\Rightarrow\semantics{S})\Leftarrow\semantics{N}}(\widetilde{likes}_{(1)})(\epsilon_{\F\semantics{N}\Rightarrow\semantics{N}}(\widetilde{his}_{(1)})(\overrightarrow{code})(\widetilde{John}_{(2)}))(\widetilde{John}_{(1)}) \otimes 
	\\
	\qquad
	\overrightarrow{does\,too}(\epsilon_{\F(\F\semantics{N}\Rightarrow\semantics{S})\Leftarrow\semantics{N}}(\widetilde{likes}_{(2)})((\epsilon_{\F\semantics{N}\Rightarrow\semantics{N}}(\widetilde{his}_{(1)})(\overrightarrow{code})(\widetilde{Bill}_{(2)})))(\epsilon_{\semantics{N}}(\widetilde{Bill}_{(2)})
	\end{array}
	\]
	Finally, we show what the strict and sloppy maps look like when specifying $\Delta$ to be $\mathbf{k}$-extension and basis copying, however we omit the $\mathbf{k}$-extension map for the sloppy reading, as this takes about 16 lines to define, and becomes highly uninformative.
	\small\[\begin{array}{l}
	\mathbf{k}\text{\bf-extension}\\
	h_{strict}(
		\widetilde{John} \otimes 
		\widetilde{likes} \otimes
		\overrightarrow{his} \otimes
		\overrightarrow{code} \otimes
		\widetilde{Bill} \otimes 
	\overrightarrow{does\,too}) :=
		\\
        \qquad\quad (likes(\ov{John}, his(\mathbf{k}, \ov{code})) \otimes \mathbf{k} (\ov{Bob}, his(\mathbf{k}, \ov{code})) +\\
        \qquad\qquad 
         \mathbf{k}(\ov{John}, his(\mathbf{k}, \ov{code}) \otimes likes(\ov{Bob}, his(\mathbf{k}, \ov{code}))) + \\
         \qquad \qquad\quad
         likes(\mathbf{k}, his(\ov{John}, \ov{code})) \otimes \mathbf{k}(\ov{Bob}, his(\ov{John}, \ov{code})) +\\ 
        \qquad \qquad \qquad \mathbf{k}(\mathbf{k}, his(\mathbf{k}, \ov{code})) \otimes likes(\ov{1}, his(\ov{John}, \ov{code}))
\\ 
\text{\bf Basis copy} \\
h_{strict}(
 \tilde{n}_{i_1} \otimes
 (\widetilde{(\widetilde{\tilde{n}_{i_2}^*\otimes s_{j_1}})\otimes n_{i_3}^*}) \otimes
 ((\widetilde{\tilde{n}_{i_4}^* \otimes n_{i_5}})\otimes n_{i_6}^*) \otimes
 n_{i_7} \otimes
 \tilde{n}_{i_8} \otimes
 ((\widetilde{\tilde{n}_{i_9}^*\otimes s_{j_2}})^*\otimes \tilde{n}_{i_{10}}^*\otimes s_{j_3})
 ) \\
:= (n_{i_2}^*(n_{i_1})s_{j_1}n_{i_3}^*(n_{i_5})n_{i_4}^*(n_{i_1})n_{i_6}^*(n_{i_7}))
\otimes \\
\qquad (n_{i_{10}}^*(n_{i_8}) s_{j_3} (n_{i_9}^* \otimes 
s_{j_2})^*(n_{i_2}^*\otimes s_{j_1})n_{i_3}^*(n_{i_5})n_{i_4}^*(n_{i_1})n_{i_6}^*(n_{i_7}))
\\
 h_{sloppy}(
 \tilde{n}_{i_1} \otimes
 (\widetilde{(\widetilde{\tilde{n}_{i_2}^*\otimes s_{j_1}})\otimes n_{i_3}^*}) \otimes
 ((\widetilde{\tilde{n}_{i_4}^* \otimes n_{i_5}})\otimes n_{i_6}^*) \otimes
 n_{i_7} \otimes
 \tilde{n}_{i_8} \otimes
 ((\widetilde{\tilde{n}_{i_9}^*\otimes s_{j_2}})^*\otimes \tilde{n}_{i_{10}}^*\otimes s_{j_3})
 ) 
 \\
:= (n_{i_2}^*(n_{i_1})s_{j_1}n_{i_3}^*(n_{i_5})n_{i_4}^*(n_{i_1})n_{i_6}^*(n_{i_7}))
\otimes \\
\qquad (n_{i_{10}}^*(n_{i_8}) s_{j_3} (n_{i_9}^* \otimes 
s_{j_2})^*(n_{i_2}^*\otimes s_{j_1})n_{i_3}^*(n_{i_5})n_{i_4}^*(n_{i_8})n_{i_6}^*(n_{i_7}))
\end{array}\] \normalsize

\section{Experiments}

We implement our copying operations on the disambiguation task of \citep{WijnSadrNAACL},  for the purpose of deciding which copying map does better in practice. The aim is to compare the performance of our linear copying maps  to each other and to the non linear copying map. We would also like to find out how well do the compositional  models do in comparison to a non compositional verb only baselines and non grammatical compositional methods such as addition. 

The disambiguation task of \citep{WijnSadrNAACL}, extends the original disambiguation task introduced in \citep{GrefenSadrEMNLP} with elliptic phrases. The original dataset of \citep{GrefenSadrEMNLP} worked with 10 ambiguous verbs and two of their meanings.  An example is the verb \emph{draw}, which is ambiguous  between \emph{depict} and \emph{pull}. The ambiguous verb  and each of its meanings are placed in subject-verb-object triples.  For the verb \emph{draw}, we have the sentences

\begin{quote}
$S$: man draw sword.\\
$S_1$: man depict sword. \\
$S_2$: man pull sword.
\end{quote}

The dataset consists of pairs of triple $(S, S_1)$  and $(S, S_2)$. The aim of the task is to build vectors for $S, S_1, S_2$, compute the cosine distances between $S, S_1$ and $S, S_2$, in order to decide which meaning of the verb is the more appropriate one in $S$. Clearly, if $S$ is closer to $S_1$, its first meaning is deemed more appropriate and if it is closer to $S_2$, its second meaning. 

In \citep{WijnSadrNAACL}, the above dataset is extended to triples with elliptic phrases. It is hypothesised  that the extended sentences provide a better base for disambiguation  and indeed  this hypothesis is verified in the paper. In the interest of space we do not go through the details of this hypothesis and the results and only provide an example. For the ambiguous verb \emph{draw},  we now work with the following sentences:

\begin{quote}
$S'$: man draw sword and artist does too.\\
$S'_1$: man depict sword and artist does too. \\
$S'_2$: man pull sword and artist does too.
\end{quote}

Vectors for each sentence with elliptic phrase are built using the procedure described in  subsection \ref{subsec:vectellipsis}, where we choose to consider vectors in Fock spaces to be of the form $(0,v,0,0,\ldots)$ for practicality. For the verb, previous experimentation has shown that cubes have a much lower performance than matrices. So although the derivational system prescribes a cube for a transitive verb, in practice it is better to approximate them by matrices. In this case, the cube contraction is approximated by  a  matrix multiplication followed by a pointwise vector multiplication. We follow previous work and  implement the  \emph{Relational}  verb matrices and  the \emph{Copy-Object} composition model. These two methods used in conjunction have consistently provided consistent good results in previous work, e.g. see \citep{GrefenSadrEMNLP,KartSadrCoNLL,milajevs-etal-2014-evaluating,mcpheat2020categorical}.  The \emph{Relational} method of building verb matrices sums the Kronecker product of the subjects and objects of the verb in the sentences across the corpus, resulting in the formula $\overline{V} := \sum_i \ov{s}_i \otimes \ov{o}_i$. Using this formula and employing a \emph{Copy-Object} method in a sentence with elliptical phrase `Sub1 Verb Obj and Sub2 does-too'   results in the following formulae for  the $\mathbf{k}$-extension copying. 
\small\begin{eqnarray*}
\mathbf{k}\mbox{-extension}:&  ((\overline{V} \times \ov{Obj}) \odot \ov{Sub1} +  (\mathbf{1} \odot \ov{Sub2})) +  ((\mathbf{1} \odot  \ov{Sub1})  + (\overline{V} \times \ov{Obj}) \odot\ov{Sub2})\\
= & ((\overline{V} \times \ov{Obj}) \odot \ov{Sub1} +  \ov{Sub2}) +  (\ov{Sub1}  + (\overline{V} \times \ov{Obj}) \odot\ov{Sub2})
\end{eqnarray*}\normalsize
Here, following previous work we are interpreting the preposition `\textit{and}' as addition, taking $k$ to be 1, and interpreting the elliptic marker `\textit{does-too}' as identity. The \emph{full} model is where the copying operation is non-linear and provides us with two proper copies of the ellipsis verb phrase, resulting in the following formula:

\small\[
\mbox{full}:  ((\overline{V} \times \ov{Obj}) \odot \ov{Sub1}) +  ((\overline{V} \times \ov{Obj}) \odot \ov{Sub2})
\]\normalsize

We use 100 dimensional pre-trained  \textbf{word2vec} and \textbf{fasttext} vectors for the $\ov{Sub1}, \ov{Sub2}, \ov{Obj}$ vectors, as well as the $\ov{s}_i$ and $\ov{o}_i$ vectors used to build our \emph{Relational} verb matrices. 

\begin{table}[h]
\centering
\begin{tabular}{lccccc}
\toprule
&&&basis & basis &\\
&{} &  full &  copy(a) &  copy(b) &  $\mathbf{k}$-extension \\
\midrule
\textbf{word2vec} &&   0.44 & 0.34 & 0.42  & 0.44 \\
\textbf{fasttext} &&   0.43 & 0.36  & 0.41 &  0.43 \\
\hline
\hline
baselines\\
\hline
\hline
\textbf{verb only} & 0.24 &&&&\\
 \textbf{additive} & 0.31 &&&&\\
 \textbf{BERT phrase}  & 0.36 &&&\\
 \hline
\hline
inter-annotator agreement& 0.58 &&&&\\
\bottomrule
\end{tabular}
\caption{Spearman's $\rho$ correlations  for the ellipsis disambiguation task;  upper bound is the inter annotator agreement score, computed in  \citep{WijnholdsThesis}. }
\label{ELLDISDis_copy_object}
\end{table}

Going through Table \ref{ELLDISDis_copy_object}, we observe that both of the \textbf{word2vec} and \textbf{fasttext} vectors provide correlations close to the upper-bound, which are slightly  higher for \textbf{word2vec}. In either case, the best performance is obtained by the full and the  $\mathbf{k}$-extension copying operations. Interestingly, the compositional models perform better than the non compositional verb-only baseline, which only provides a correlation of 0.24. This low correlation is improved to 0.31 when a non grammatical additive base line is used and to reaches its maximum when 0.35 when pre-trained \textbf{BERT}  vectors are used. It is worth noting that  the compositional state of the art of the Ellipsis Disambiguation dataset is 0.58, as reported in \citep{wijnholds-etal-2020} and  is obtained via neural verb matrices. In the same paper, it is explained how fine-tuning a \textbf{BERT} model provides the best non compositional performance of 0.65. Given that the performances reported here are obtained via pre-trained vectors, we expect that fine tuning our vectors or matrices will provide better results than those reported here. We also expect that the use of neural verb matrices in our linear copying operations will improve the results.

\section{Conclusion and Future Work}
Following the style of \citep{Coeckeetal2013}, where the authors develop a functorial vector space semantics and string diagrams  for Lambek calculus, in another paper \citep{mcpheat2020categorical} we  developed a functorial categorical semantics for Lambek calculus with a Relevant Modality of  \citep{Kanovich2016}.  This logic extends Lambek calculus with a relevant modality that allows for limited contraction and permutation. The motivation for development of $\blstar$ is  the use of limited contraction and permutation  to reason about  parasitic gaps in  line of research initiated  in \citep{Morrilletal1990} and followed up on in  \citep{morrill2016logic} and \citep{Morrill2017,Morrill2018}. In  this paper, we do not go through the categorical model, as it is extensive and  constitutes the contribution of  another paper. Instead, we show how one can assign a vector space semantics to this calculus directly, without going through category theory,  and with the help of the notion of tensor algebras and a particular finite kind used in Quantum Mechanics: Fermionic Fock Spaces.  Inspired by the work of \citep{jager1998multi,jager2006anaphora}  and \citep{WijnSadr2018,WijnSadr2019}, for the first time, we apply the Lambek calculus with the Relevant Modality to reason about anaphora with ellipsis and develop closed form linear algebraic terms for the results of the corresponding derivations. We experiment with our model  on the Ellipsis Disambiguation dataset of \citep{WijnSadrNAACL}  and observed that our $\mathbf{k}$-extension  linear copying operation  provides the same results as a full non linear copying operation. These models  significantly outperform the non compositional and non grammatical baselines.  They are, however,  improved by a fine tuned \textbf{bert} model and also in a compositional model with neural verb matrices \citep{wijnholds-etal-2020}.
Our other contribution  is that we show how this vector space semantics is able to distinguish between the two readings of the ambiguous anaphora with ellipsis cases. Indeed and as desired and aligned with the standard literature on coreference modelling \citep{KentEllipsis}, we obtain two different  linear maps as semantics of these cases, one for the strict reading and one for the sloppy reading.  This overcomes the weakness of  previous work \citep{WijnSadr2019}, where a Lambek calculus with a copying modality was used to model coreference.

\bibliographystyle{jcs}
\bibliography{references}
\end{document}

%% file: tikzpreamble.tex
	\usepackage{tikz}
\usepackage{pgfplots}
\usetikzlibrary{trees}
\usetikzlibrary{topaths}
\usetikzlibrary{decorations.pathmorphing}
\usetikzlibrary{decorations.markings}
\usetikzlibrary{matrix,backgrounds,folding}
\usetikzlibrary{chains,scopes,positioning,fit}
\usetikzlibrary{arrows,shadows}
\usetikzlibrary{calc} 
\usetikzlibrary{chains}
\usetikzlibrary{shapes,shapes.geometric,shapes.misc}

\tikzset{->-/.style={decoration={
  markings,
  mark=at position .5 with {\arrow{>}}},postaction={decorate}}}

\pgfdeclarelayer{edgelayer}
\pgfdeclarelayer{nodelayer}
\pgfsetlayers{background,edgelayer,nodelayer,main}

\tikzstyle{new style 0}=[fill=white, draw=black, shape=rectangle, font=\scriptsize]
\tikzstyle{new style 1}=[fill=white, draw=black, shape=circle, font=\tiny]
\tikzstyle{none}=[fill=none, font=\tiny]
\tikzstyle{lab}=[fill=white, font=\tiny]
\tikzstyle{dot}=[circle,fill=black,draw=black]
\tikzstyle{counit}=[fill=black, shape=circle]
\tikzstyle{sq}=[rectangle, fill=white, draw=black]
\tikzstyle{bbox1}=[fill=white, draw=black, shape=rectangle, minimum width=4.5cm]
\tikzstyle{bbox2}=[fill=white, draw=black, shape=rectangle, minimum width=2cm]
\tikzstyle{every picture}=[baseline=(current bounding box).east,scale=0.5,node distance=5mm]
\tikzstyle{every loop}=[]

\tikzstyle{downArrow}=[->-, line width = 1pt]
\tikzstyle{thickArr}=[->-, line width = 2pt]
\tikzstyle{thickerArr}=[->-, line width = 4pt]

\tikzstyle{(null)}=[]
\tikzstyle{plain}=[]

%% file: obsHomsComp.tikz
\begin{tikzpicture}
	\begin{pgfonlayer}{nodelayer}
		\node [style=none] (0) at (-4.25, 5) {};
		\node [style=none] (1) at (-4.25, 0) {};
		\node [style=none] (2) at (-3.75, 2.5) {$V$};
		\node [style=none] (9) at (0, 5) {};
		\node [style=none] (10) at (0, 0) {};
		\node [style=none] (11) at (4, 5) {};
		\node [style=none] (14) at (4, 0) {};
		\node [style=new style 0] (15) at (0, 2.5) {$f$};
		\node [style=new style 0] (16) at (4, 3.5) {$f$};
		\node [style=new style 0] (17) at (4, 1.5) {$g$};
		\node [style=none] (18) at (4.75, 4.5) {$V$};
		\node [style=none] (19) at (4.75, 2.5) {$W$};
		\node [style=none] (20) at (4.75, 0.5) {$U$};
		\node [style=none] (21) at (1, 4) {$V$};
		\node [style=none] (22) at (1, 1) {$W$};
	\end{pgfonlayer}
	\begin{pgfonlayer}{edgelayer}
		\draw [style=downArrow] (0.center) to (1.center);
		\draw [style=downArrow] (9.center) to (15);
		\draw [style=downArrow] (15) to (10.center);
		\draw [style=downArrow] (11.center) to (16);
		\draw [style=downArrow] (16) to (17);
		\draw [style=downArrow] (17) to (14.center);
	\end{pgfonlayer}
\end{tikzpicture}

%% file: monObsHoms.tikz
\begin{tikzpicture}
	\begin{pgfonlayer}{nodelayer}
		\node [style=none] (0) at (-4.25, 5) {};
		\node [style=none] (1) at (-4.25, 0) {};
		\node [style=none] (2) at (-2.25, 5) {};
		\node [style=none] (3) at (-2.25, 0) {};
		\node [style=none] (4) at (-3.75, 2.5) {$V$};
		\node [style=none] (5) at (-1.75, 2.5) {$W$};
		\node [style=none] (6) at (6.75, 5) {};
		\node [style=none] (7) at (9, 0) {};
		\node [style=new style 0] (8) at (6.75, 2.5) {$f$};
		\node [style=new style 0] (9) at (9, 2.5) {$g$};
		\node [style=none] (10) at (7.5, 4) {$V$};
		\node [style=none] (11) at (7.5, 1.5) {$W$};
		\node [style=none] (12) at (10, 1.5) {$W'$};
		\node [style=none] (13) at (9, 5) {};
		\node [style=none] (14) at (6.75, 0) {};
		\node [style=none] (15) at (10, 4) {$V'$};
		\node [style=none] (16) at (-8, 5) {};
		\node [style=none] (17) at (-8, 0) {};
		\node [style=none] (18) at (-6.75, 2.5) {$V \otimes W$};
		\node [style=none] (19) at (-5.25, 2.5) {$=$};
		\node [style=none] (20) at (3, 5) {};
		\node [style=new style 0] (21) at (3, 2.5) {$f \otimes g$};
		\node [style=none] (22) at (4.25, 4) {$V \otimes V'$};
		\node [style=none] (23) at (4.25, 1.5) {$W \otimes W'$};
		\node [style=none] (24) at (3, 0) {};
		\node [style=none] (25) at (5.5, 2.5) {$=$};
	\end{pgfonlayer}
	\begin{pgfonlayer}{edgelayer}
		\draw [style=downArrow] (0.center) to (1.center);
		\draw [style=downArrow] (2.center) to (3.center);
		\draw [style=downArrow] (6.center) to (8);
		\draw [style=downArrow] (9) to (7.center);
		\draw [style=downArrow] (13.center) to (9);
		\draw [style=downArrow] (8) to (14.center);
		\draw [style=downArrow] (16.center) to (17.center);
		\draw [style=downArrow] (20.center) to (21);
		\draw [style=downArrow] (21) to (24.center);
	\end{pgfonlayer}
\end{tikzpicture}

%% file: clasps.tikz
\begin{tikzpicture}
	\begin{pgfonlayer}{nodelayer}
		\node [style=none] (0) at (6, 1.25) {};
		\node [style=none] (1) at (8, 1.25) {};
		\node [style=none] (2) at (6, -2) {};
		\node [style=none] (3) at (8, -2) {};
		\node [style=none] (4) at (8, -1) {};
		\node [style=new style 1] (5) at (6, -1) {};
		\node [style=none] (6) at (-2.25, 1.25) {};
		\node [style=none] (7) at (-4.25, 1.25) {};
		\node [style=none] (8) at (-2.25, -2) {};
		\node [style=none] (9) at (-4.25, -2) {};
		\node [style=none] (10) at (-4.25, -1) {};
		\node [style=new style 1] (11) at (-2.25, -1) {};
		\node [style=none] (12) at (-7.75, 1.25) {};
		\node [style=none] (13) at (-7.75, -2) {};
		\node [style=none] (14) at (2, 1.25) {};
		\node [style=none] (15) at (2, -2) {};
		\node [style=none] (16) at (5, -0.75) {$=$};
		\node [style=none] (17) at (-5.25, -0.75) {$=$};
		\node [style=none] (18) at (3, 0.5) {$W\Leftarrow V$};
		\node [style=none] (19) at (6.5, 0.5) {$W$};
		\node [style=none] (20) at (8.5, 0.5) {$V$};
		\node [style=none] (21) at (-6.75, 0.5) {$V \Rightarrow W$};
		\node [style=none] (22) at (-3.75, 0.5) {$V$};
		\node [style=none] (23) at (-1.5, 0.5) {$W$};
	\end{pgfonlayer}
	\begin{pgfonlayer}{edgelayer}
		\draw (4.center) to (5);
		\draw [style=downArrow] (3.center) to (1.center);
		\draw [style=downArrow] (0.center) to (2.center);
		\draw (10.center) to (11);
		\draw [style=downArrow] (9.center) to (7.center);
		\draw [style=downArrow] (6.center) to (8.center);
		\draw [style=downArrow] (14.center) to (15.center);
		\draw [style=downArrow] (12.center) to (13.center);
	\end{pgfonlayer}
\end{tikzpicture}

%% file: cups.tikz
\begin{tikzpicture}
	\begin{pgfonlayer}{nodelayer}
		\node [style=none] (0) at (2.25, 2.25) {};
		\node [style=none] (1) at (4.5, 2.25) {};
		\node [style=none] (2) at (7, 2.25) {};
		\node [style=none] (3) at (2.25, -0.75) {};
		\node [style=none] (4) at (4.5, -0.75) {};
		\node [style=none] (5) at (7, -2.5) {};
		\node [style=none] (6) at (2.75, 1.75) {$V$};
		\node [style=none] (7) at (5.25, 1.75) {$V$};
		\node [style=none] (8) at (7.75, 1.75) {$W$};
		\node [style=new style 1] (9) at (7, 0) {};
		\node [style=none] (10) at (4.5, 0) {};
		\node [style=none] (11) at (-7.5, 2) {};
		\node [style=none] (12) at (-5, 2) {};
		\node [style=none] (13) at (-2.5, 2) {};
		\node [style=none] (14) at (-5.25, -0.75) {};
		\node [style=none] (15) at (-5, -0.75) {};
		\node [style=none] (16) at (-4.5, -0.75) {};
		\node [style=none] (17) at (-6.75, 1.5) {$V$};
		\node [style=none] (18) at (-4, 1.5) {$V$};
		\node [style=none] (19) at (-1.75, 1.5) {$W$};
		\node [style=new style 1] (20) at (-4, 0.25) {};
		\node [style=none] (21) at (-5, 0.25) {};
		\node [style=new style 0] (22) at (-5, -1) {$ev_{V,W}$};
		\node [style=none] (23) at (0, 0) {$=$};
		\node [style=none] (24) at (-5, -2.5) {};
		\node [style=none] (25) at (-4, -2.5) {$W$};
	\end{pgfonlayer}
	\begin{pgfonlayer}{edgelayer}
		\draw (10.center) to (9);
		\draw [style=downArrow] (4.center) to (1.center);
		\draw [style=downArrow] (2.center) to (5.center);
		\draw [style=downArrow] (0.center) to (3.center);
		\draw [style=downArrow, bend right=90, looseness=1.25] (3.center) to (4.center);
		\draw (21.center) to (20);
		\draw [style=downArrow] (15.center) to (12.center);
		\draw [style=downArrow, in=90, out=-90] (13.center) to (16.center);
		\draw [style=downArrow, in=90, out=-90, looseness=1.25] (11.center) to (14.center);
		\draw [style=downArrow] (22) to (24.center);
	\end{pgfonlayer}
\end{tikzpicture}

%% file: vdots.tikz
\begin{tikzpicture}
	\begin{pgfonlayer}{nodelayer}
		\node [style=none] (0) at (-3, 2) {};
		\node [style=none] (1) at (-3, 0) {};
		\node [style=none] (2) at (-3, -0.25) {$\vdots$};
		\node [style=none] (3) at (-4, -0.5) {};
		\node [style=none] (4) at (-2, -0.5) {};
		\node [style=none] (5) at (-4, -2.75) {};
		\node [style=none] (6) at (-2, -2.75) {};
		\node [style=none] (7) at (-4, -2) {};
		\node [style=new style 1] (8) at (-2, -2) {};
		\node [style=none] (9) at (-1.5, 1.5) {$V\Rightarrow W$};
		\node [style=none] (10) at (-3.5, -1.25) {$V$};
		\node [style=none] (11) at (-1.25, -1.25) {$W$};
		\node [style=none] (12) at (3, 2) {};
		\node [style=none] (13) at (3, 0) {};
		\node [style=none] (14) at (3, -0.25) {$\vdots$};
		\node [style=none] (15) at (4, -0.5) {};
		\node [style=none] (16) at (2, -0.5) {};
		\node [style=none] (17) at (4, -2.75) {};
		\node [style=none] (18) at (2, -2.75) {};
		\node [style=none] (19) at (4, -2) {};
		\node [style=new style 1] (20) at (2, -2) {};
		\node [style=none] (21) at (4.5, 1.5) {$W \Leftarrow V$};
		\node [style=none] (22) at (2.5, -1.25) {$W$};
		\node [style=none] (23) at (4.5, -1.25) {$V$};
	\end{pgfonlayer}
	\begin{pgfonlayer}{edgelayer}
		\draw [style=downArrow] (0.center) to (1.center);
		\draw [style=downArrow] (5.center) to (3.center);
		\draw [style=downArrow] (4.center) to (6.center);
		\draw (7.center) to (8);
		\draw [style=downArrow] (12.center) to (13.center);
		\draw [style=downArrow] (17.center) to (15.center);
		\draw [style=downArrow] (16.center) to (18.center);
		\draw (19.center) to (20);
	\end{pgfonlayer}
\end{tikzpicture}

%% file: backslashL.tikz
\begin{tikzpicture}
	\begin{pgfonlayer}{nodelayer}
		\node [style=none] (0) at (3, 0.75) {$\semantics{B}$};
		\node [style=none] (1) at (1, 0.75) {$\semantics{A}$};
		\node [style=none] (2) at (-4, 1.5) {};
		\node [style=none] (3) at (-2, 1.5) {};
		\node [style=none] (4) at (4.25, 1.5) {};
		\node [style=none] (5) at (2.25, 1.5) {};
		\node [style=none] (6) at (0.25, 1.5) {};
		\node [style=none] (7) at (2.25, -0.75) {};
		\node [style=none] (8) at (0.25, -0.75) {};
		\node [style=none] (9) at (0.25, -0.5) {};
		\node [style=new style 1] (10) at (2.25, -0.5) {};
		\node [style=none] (11) at (-4, -0.75) {};
		\node [style=none] (12) at (4.25, -0.75) {};
		\node [style=none] (13) at (-3.25, 0.75) {$\semantics{\Delta_1}$};
		\node [style=none] (14) at (-1.25, 0.75) {$\semantics{\Gamma}$};
		\node [style=none] (15) at (5, 0.75) {$\semantics{\Delta_2}$};
		\node [style=new style 0] (16) at (-2, -0.5) {$f$};
		\node [style=none] (17) at (-2, -0.75) {};
		\node [style=new style 0,minimum width=1.2cm] (18) at (0, -3.5) {$g$};
		\node [style=none] (19) at (0, -5.75) {};
		\node [style=none] (20) at (1, -4.75) {$\semantics{C}$};
		\node [style=none] (21) at (-0.5, -3.25) {};
		\node [style=none] (22) at (0, -3.25) {};
		\node [style=none] (23) at (0.5, -3.25) {};
	\end{pgfonlayer}
	\begin{pgfonlayer}{edgelayer}
		\draw (9.center) to (10);
		\draw [style=downArrow] (8.center) to (6.center);
		\draw [style=downArrow] (5.center) to (7.center);
		\draw [style=downArrow] (2.center) to (11.center);
		\draw [style=downArrow] (4.center) to (12.center);
		\draw [style=downArrow] (3.center) to (16);
		\draw [style=downArrow] (16) to (17.center);
		\draw [style=downArrow, bend right=90, looseness=1.25] (17.center) to (8.center);
		\draw [style=downArrow] (18) to (19.center);
		\draw [style=downArrow, in=90, out=-90, looseness=0.75] (12.center) to (23.center);
		\draw [style=downArrow, in=90, out=-90] (7.center) to (22.center);
		\draw [style=downArrow, in=90, out=-90] (11.center) to (21.center);
	\end{pgfonlayer}
\end{tikzpicture}

%% file: perm2.tikz
\begin{tikzpicture}
	\begin{pgfonlayer}{nodelayer}
		\node [style=none] (0) at (-3.25, 2.5) {};
		\node [style=none] (1) at (-1, 2.5) {};
		\node [style=none] (2) at (1, 2.5) {};
		\node [style=none] (3) at (3.25, 2.5) {};
		\node [style=none] (4) at (-2.5, 2.25) {$\semantics{\Delta_1}$};
		\node [style=none] (5) at (-0.25, 2.25) {$\semantics{\Gamma}$};
		\node [style=none] (6) at (1.75, 2.25) {$\F\semantics{A}$};
		\node [style=none] (7) at (4, 2.25) {$\semantics{\Delta_2}$};
		\node [style=none] (8) at (-1, 0.25) {};
		\node [style=none] (9) at (-0.5, 0.25) {};
		\node [style=none] (10) at (0.5, 0.25) {};
		\node [style=none] (11) at (1, 0.25) {};
		\node [style=new style 0, minimum width=1.3cm] (12) at (0, 0.25) {$f$};
		\node [style=none] (13) at (0, -1.75) {};
		\node [style=none] (14) at (0.5, -1.25) {$\semantics{B}$};
		\node [style=none] (15) at (0, 1.25) {};
	\end{pgfonlayer}
	\begin{pgfonlayer}{edgelayer}
		\draw [style=downArrow] (12) to (13.center);
		\draw [style=downArrow, in=90, out=-90] (0.center) to (8.center);
		\draw [style=downArrow, in=90, out=-90] (3.center) to (11.center);
		\draw [style=downArrow, in=135, out=-90, looseness=1.25] (1.center) to (15.center);
		\draw [style=downArrow, in=45, out=-90, looseness=1.25] (2.center) to (15.center);
		\draw [style=downArrow, in=90, out=-135] (15.center) to (9.center);
		\draw [style=downArrow, in=90, out=-45] (15.center) to (10.center);
	\end{pgfonlayer}
\end{tikzpicture}

%% file: contr2.tikz
\begin{tikzpicture}
	\begin{pgfonlayer}{nodelayer}
		\node [style=none] (0) at (0, 1.25) {};
		\node [style=none] (1) at (-0.75, -3) {};
		\node [style=none] (2) at (0.75, -3) {};
		\node [style=new style 0] (3) at (0, -0.75) {$\Delta_{\semantics{A}}$};
		\node [style=none] (4) at (0.75, 0.75) {$\semantics{A}$};
		\node [style=none] (5) at (-0.25, -2.25) {$\semantics{A}$};
		\node [style=none] (6) at (1.25, -2.25) {$\semantics{A}$};
		\node [style=new style 0, minimum width=1.7cm] (7) at (0, -3) {$f$};
		\node [style=none] (8) at (-1.25, 0.5) {$\semantics{\Delta_1}$};
		\node [style=none] (9) at (3, 0.5) {$\semantics{\Delta_2}$};
		\node [style=none] (10) at (2, 1.25) {};
		\node [style=none] (11) at (-2, 1.25) {};
		\node [style=none] (12) at (-1.5, -3) {};
		\node [style=none] (13) at (1.5, -3) {};
		\node [style=none] (14) at (0, -4.5) {};
		\node [style=none] (15) at (0.75, -4) {$\semantics{B}$};
	\end{pgfonlayer}
	\begin{pgfonlayer}{edgelayer}
		\draw [style=thickArr] (0.center) to (3);
		\draw [style=thickArr, in=90, out=-135, looseness=0.75] (3) to (1.center);
		\draw [style=thickArr, in=90, out=-45, looseness=0.75] (3) to (2.center);
		\draw [style=downArrow, in=90, out=-90, looseness=0.25] (11.center) to (12.center);
		\draw [style=downArrow, in=90, out=-90, looseness=0.75] (10.center) to (13.center);
		\draw [style=downArrow] (7) to (14.center);
	\end{pgfonlayer}
\end{tikzpicture}

%% file: JohnSleeps1.tikz
\begin{tikzpicture}
	\begin{pgfonlayer}{nodelayer}
		\node [style=none] (0) at (-3, 2.75) {};
		\node [style=new style 0] (4) at (-3, 3.5) {John};
		\node [style=new style 0] (5) at (0.25, 3.5) {sleeps};
		\node [style=new style 0] (6) at (4, 3.5) {He};
		\node [style=new style 0] (7) at (7.25, 3.5) {snores};
		\node [style=none] (8) at (-2.25, 2.25) {$\semantics{!N}$};
		\node [style=none] (9) at (0, 2.25) {$\semantics{N}$};
		\node [style=none] (13) at (-0.5, 2.75) {};
		\node [style=none] (14) at (1, 2.75) {};
		\node [style=none] (15) at (3.25, 2.75) {};
		\node [style=new style 0] (16) at (-3, 1.5) {$\Delta_{\semantics{N}}$};
		\node [style=none] (19) at (4.75, 2.75) {};
		\node [style=none] (20) at (6.5, 2.75) {};
		\node [style=none] (21) at (8, 2.75) {};
		\node [style=none] (22) at (-0.5, 0) {};
		\node [style=none] (23) at (1, 0) {};
		\node [style=none] (24) at (3.25, -0.5) {};
		\node [style=none] (25) at (4.75, -1.25) {};
		\node [style=none] (26) at (6.5, -1.25) {};
		\node [style=none] (27) at (8, 0) {};
		\node [style=none] (28) at (-0.5, 1.75) {};
		\node [style=none] (29) at (3.25, 1.75) {};
		\node [style=none] (30) at (6.5, 1.75) {};
		\node [style=new style 1] (31) at (1, 1.75) {};
		\node [style=new style 1] (32) at (4.75, 1.75) {};
		\node [style=new style 1] (33) at (8, 1.75) {};
		\node [style=none] (35) at (1.75, -0.5) {};
		\node [style=new style 0] (36) at (-4, 0) {$\epsilon_{\semantics{N}}$};
		\node [style=none] (37) at (1, -2.5) {};
		\node [style=none] (38) at (8, -2.5) {};
		\node [style=none] (39) at (-3.75, -1.25) {$\semantics{N}$};
		\node [style=none] (40) at (2, -2) {$\semantics{S}$};
		\node [style=none] (41) at (8.5, -2) {$\semantics{S}$};
		\node [style=none] (42) at (1.5, 2.25) {$\semantics{S}$};
		\node [style=none] (43) at (7.075, 2.25) {$\semantics{N}$};
		\node [style=none] (44) at (8.575, 2.25) {$\semantics{S}$};
		\node [style=none] (45) at (3.825, 2.25) {$\semantics{!N}$};
		\node [style=none] (46) at (5.325, 2.25) {$\semantics{N}$};
	\end{pgfonlayer}
	\begin{pgfonlayer}{edgelayer}
		\draw [style=downArrow] (0.center) to (16);
		\draw [style=downArrow] (22.center) to (13.center);
		\draw [style=downArrow] (14.center) to (23.center);
		\draw [style=downArrow] (24.center) to (15.center);
		\draw [style=downArrow] (19.center) to (25.center);
		\draw [style=downArrow] (26.center) to (20.center);
		\draw [style=downArrow] (21.center) to (27.center);
		\draw (28.center) to (31);
		\draw (29.center) to (32);
		\draw (30.center) to (33);
		\draw [style=downArrow, bend right=90, looseness=1.25] (25.center) to (26.center);
		\draw [style=downArrow, bend right=90, looseness=1.75] (35.center) to (24.center);
		\draw [style=downArrow, bend right=90] (36) to (22.center);
		\draw [style=downArrow] (23.center) to (37.center);
		\draw [style=downArrow] (27.center) to (38.center);
		\draw [style=downArrow, in=90, out=-45] (16) to (35.center);
		\draw [style=downArrow, in=90, out=-135] (16) to (36);
	\end{pgfonlayer}
\end{tikzpicture}

%% file: ellipsis.tikz
\begin{tikzpicture}
	\begin{pgfonlayer}{nodelayer}
		\node [style=none] (0) at (-7.5, 6) {};
		\node [style=none] (2) at (-4, 6) {};
		\node [style=none] (3) at (0, 6) {};
		\node [style=none] (4) at (2, 6) {};
		\node [style=none] (5) at (4, 6) {};
		\node [style=none] (6) at (6, 6) {};
		\node [style=none] (7) at (10.75, 6) {};
		\node [style=new style 0] (8) at (-7.5, 6.5) {John};
		\node [style=new style 0] (9) at (-2.5, 6.5) {plays};
		\node [style=new style 0] (10) at (2, 6.5) {guitar};
		\node [style=new style 0] (11) at (4, 6.5) {Mary};
		\node [style=new style 0] (12) at (8.25, 6.5) {does too};
		\node [style=none] (13) at (-6.75, 5.5) {$\semantics{N}$};
		\node [style=none] (15) at (-2.25, 5.5) {$\F(\semantics{N}\Rightarrow \semantics{S})$};
		\node [style=none] (19) at (7.75, 5.5) {$\F(\semantics{N}\Rightarrow \semantics{S})$};
		\node [style=none] (20) at (12.25, 5.5) {$\semantics{N}\Rightarrow \semantics{S}$};
		\node [style=none] (21) at (0.75, 5.5) {$\semantics{N}$};
		\node [style=none] (22) at (2.5, 5.5) {$\semantics{N}$};
		\node [style=none] (23) at (4.5, 5.5) {$\semantics{N}$};
		\node [style=none] (24) at (-7.5, -1.5) {};
		\node [style=none] (26) at (0, 4.25) {};
		\node [style=none] (27) at (2, 4.25) {};
		\node [style=none] (28) at (4, -0.5) {};
		\node [style=none] (29) at (6, 4.25) {};
		\node [style=none] (30) at (10.75, 4.25) {};
		\node [style=new style 1] (31) at (-4, 5) {};
		\node [style=new style 1] (32) at (10.75, 5) {};
		\node [style=none] (33) at (6, 5) {};
		\node [style=none] (34) at (0, 5) {};
		\node [style=new style 0] (35) at (-4, 3.25) {$\Delta_{\semantics{N}\Rightarrow \semantics{S}}$};
		\node [style=none] (37) at (-2.75, 1.75) {};
		\node [style=none] (38) at (6, 0.25) {};
		\node [style=none] (39) at (4.75, 0.25) {};
		\node [style=none] (40) at (10.75, 2) {};
		\node [style=none] (41) at (10.75, 1.5) {$\vdots$};
		\node [style=none] (42) at (9.75, 1.25) {};
		\node [style=none] (43) at (11.75, 1.25) {};
		\node [style=none] (44) at (9.75, -0.5) {};
		\node [style=none] (45) at (11.75, -0.5) {};
		\node [style=new style 1] (46) at (11.75, 0) {};
		\node [style=none] (47) at (9.75, 0) {};
		\node [style=none] (48) at (10.5, 0.75) {$\semantics{N}$};
		\node [style=none] (50) at (11.75, -2.5) {};
		\node [style=none] (51) at (-5.25, 0.5) {};
		\node [style=none] (52) at (-5.25, 0) {$\vdots$};
		\node [style=none] (53) at (-6.25, -0.25) {};
		\node [style=none] (54) at (-4.25, -0.25) {};
		\node [style=none] (55) at (-5.5, -0.5) {$\semantics{N}$};
		\node [style=none] (56) at (-3.5, -0.5) {$\semantics{S}$};
		\node [style=none] (59) at (-6.25, -1.5) {};
		\node [style=none] (60) at (-7.5, -1.5) {};
		\node [style=none] (61) at (-4.25, -2.5) {};
		\node [style=new style 1] (62) at (-4.25, -1.25) {};
		\node [style=none] (63) at (-6.25, -1.25) {};
		\node [style=none] (64) at (12.5, 0.75) {$\semantics{S}$};
		\node [style=new style 0] (65) at (-5.25, 1.5) {$\epsilon_{\semantics{N}\Rightarrow \semantics{S}}$};
	\end{pgfonlayer}
	\begin{pgfonlayer}{edgelayer}
		\draw (31) to (34.center);
		\draw (33.center) to (32);
		\draw [style=downArrow] (7.center) to (30.center);
		\draw [style=downArrow] (26.center) to (3.center);
		\draw [style=downArrow] (4.center) to (27.center);
		\draw [style=downArrow] (5.center) to (28.center);
		\draw [style=downArrow] (0.center) to (24.center);
		\draw [style=downArrow] (29.center) to (6.center);
		\draw [style=downArrow, bend left=90, looseness=1.50] (27.center) to (26.center);
		\draw [style=downArrow, in=90, out=-75] (35) to (37.center);
		\draw [style=downArrow, in=90, out=-90, looseness=0.50] (37.center) to (39.center);
		\draw [style=downArrow, bend right=90, looseness=1.25] (39.center) to (38.center);
		\draw [style=downArrow] (38.center) to (29.center);
		\draw [style=downArrow] (30.center) to (40.center);
		\draw (47.center) to (46);
		\draw [style=downArrow, bend right=90, looseness=1.25] (28.center) to (44.center);
		\draw [style=downArrow] (43.center) to (45.center);
		\draw [style=downArrow] (45.center) to (50.center);
		\draw [style=downArrow] (44.center) to (42.center);
		\draw [style=downArrow, bend right=90, looseness=1.50] (60.center) to (59.center);
		\draw [style=downArrow] (59.center) to (53.center);
		\draw [style=downArrow] (54.center) to (61.center);
		\draw (63.center) to (62);
		\draw [style=downArrow] (65) to (51.center);
		\draw [style=downArrow] (2.center) to (35);
		\draw [style=downArrow, in=90, out=-105] (35) to (65);
	\end{pgfonlayer}
\end{tikzpicture}

%% file: JohnLikesHisCode.tikz
\begin{tikzpicture}
	\begin{pgfonlayer}{nodelayer}
		\node [style=none] (0) at (-8, 3.5) {};
		\node [style=none] (1) at (-4.5, 3.5) {};
		\node [style=none] (2) at (0.5, 3.5) {};
		\node [style=none] (3) at (3.5, 3.5) {};
		\node [style=none] (4) at (5.5, 3.5) {};
		\node [style=none] (5) at (7.5, 3.5) {};
		\node [style=none] (6) at (9.5, 3.5) {};
		\node [style=none] (7) at (13.5, 3.5) {};
		\node [style=new style 0] (8) at (-8, 4.5) {John};
		\node [style=new style 0] (9) at (-4.5, 4.5) {likes};
		\node [style=new style 0] (10) at (2, 4.5) {his};
		\node [style=new style 0] (11) at (5.5, 4.5) {code};
		\node [style=new style 0] (12) at (7.5, 4.5) {Bill};
		\node [style=new style 0] (13) at (11, 4.5) {does too};
		\node [style=none] (14) at (-9.25, 3.25) {$\F\semantics{N}$};
		\node [style=lab] (15) at (-3.25, 3.25) {$\F((\F(\F\semantics{N} \Rightarrow \semantics{S}))\Leftarrow \semantics{N})$};
		\node [style=lab] (16) at (1.5, 3.25) {$\F(\semantics{!N}\Rightarrow \semantics{N}) $};
		\node [style=none] (17) at (4, 3.25) {$\semantics{N}$};
		\node [style=none] (18) at (6.05, 3.25) {$\semantics{N}$};
		\node [style=none] (19) at (8.05, 3.25) {$\F\semantics{N}$};
		\node [style=none] (20) at (11.1, 3.25) {$\F(\F\semantics{N} \Rightarrow \semantics{S})$};
		\node [style=none] (21) at (14.85, 3.25) {$\F\semantics{N} \Rightarrow \semantics{S}$};
		\node [style=none] (25) at (3.5, 1) {};
		\node [style=none] (26) at (5.5, 1) {};
		\node [style=none] (27) at (7.5, 1) {};
		\node [style=none] (28) at (9.5, 1) {};
		\node [style=none] (29) at (13.5, 0.75) {};
		\node [style=none] (30) at (9.5, 2) {};
		\node [style=none] (31) at (3.5, 2) {};
		\node [style=new style 1] (32) at (13.5, 2) {};
		\node [style=new style 1] (33) at (0.5, 2) {};
		\node [style=new style 0] (34) at (-4.5, -0.25) {$\epsilon_{(\F(\F\semantics{N}\Rightarrow S))\Leftarrow \semantics{N}}$};
		\node [style=new style 0] (35) at (0.5, 1) {$\epsilon_{ \F\semantics{N} \Rightarrow \semantics{N} }$};
		\node [style=new style 0] (36) at (-8, -0.25) {$\Delta_{\semantics{N}}$};
		\node [style=none] (37) at (-0.25, -0.75) {};
		\node [style=none] (38) at (1.25, -0.75) {};
		\node [style=none] (39) at (0.5, -0.25) {$\vdots$};
		\node [style=none] (40) at (-0.25, -2) {};
		\node [style=none] (42) at (-0.25, -1.5) {};
		\node [style=new style 1] (43) at (1.25, -1.5) {};
		\node [style=none] (44) at (-9.25, -2.5) {};
		\node [style=none] (47) at (-8.5, -2.25) {$\F\semantics{N}$};
		\node [style=none] (48) at (-5.75, -1.75) {$\F\semantics{N}$};
		\node [style=none] (50) at (-4.5, -3.5) {};
		\node [style=none] (51) at (-5.5, -4.5) {};
		\node [style=none] (52) at (-3.5, -4.5) {};
		\node [style=none] (53) at (-4.5, -3.75) {$\vdots$};
		\node [style=none] (55) at (-3.5, -6) {};
		\node [style=none] (56) at (-3.5, -5.25) {};
		\node [style=new style 1] (57) at (-5.5, -5.25) {};
		\node [style=none] (58) at (-7.075, -4.75) {$\F(\F\semantics{N}\Rightarrow \semantics{S})$};
		\node [style=none] (59) at (-3, -5) {$\semantics{N}$};
		\node [style=none] (60) at (-1.25, -2) {$\F\semantics{N}$};
		\node [style=none] (61) at (2, -2) {$\semantics{N}$};
		\node [style=none] (62) at (1.25, -6) {};
		\node [style=new style 0] (63) at (-5.5, -6.5) {$\Delta_{\F\semantics{N} \Rightarrow \semantics{S}}$};
		\node [style=new style 0] (67) at (-6.25, -8.5) {$\epsilon_{\F\semantics{N}\Rightarrow \semantics{S}}$};
		\node [style=none] (68) at (-6.25, -9.5) {};
		\node [style=none] (69) at (-7, -10.25) {};
		\node [style=none] (70) at (-5.5, -10.25) {};
		\node [style=none] (71) at (-6.25, -9.75) {$\vdots$};
		\node [style=none] (72) at (-7, -11.75) {};
		\node [style=none] (73) at (-5.5, -12.75) {};
		\node [style=none] (74) at (-7, -10.75) {};
		\node [style=new style 1] (75) at (-5.5, -10.75) {};
		\node [style=none] (76) at (-7.75, -11.25) {$\F\semantics{N}$};
		\node [style=none] (77) at (-4.75, -12.5) {$\semantics{S}$};
		\node [style=none] (78) at (-9.25, -11.75) {};
		\node [style=none] (79) at (9.5, -10.5) {};
		\node [style=none] (80) at (6.75, -10.5) {};
		\node [style=none] (81) at (12.5, 0) {};
		\node [style=none] (82) at (14.5, 0) {};
		\node [style=none] (83) at (13.5, 0.5) {$\vdots$};
		\node [style=none] (84) at (12.5, -1.75) {};
		\node [style=none] (85) at (14.5, -1.75) {};
		\node [style=none] (86) at (12.5, -0.75) {};
		\node [style=new style 1] (87) at (14.5, -0.75) {};
		\node [style=none] (88) at (11.925, -1.75) {$\F\semantics{N}$};
		\node [style=none] (89) at (15, -1.75) {$\semantics{S}$};
		\node [style=none] (90) at (7.5, -2) {};
		\node [style=none] (91) at (14.5, -9.5) {};
		\node [style=none] (92) at (14.5, -13) {};
		\node [style=none] (93) at (15, -12.75) {$\semantics{S}$};
		\node [style=none] (94) at (0.5, 0) {};
		\node [style=none] (95) at (-2, -2.25) {};
	\end{pgfonlayer}
	\begin{pgfonlayer}{edgelayer}
		\draw [style=downArrow] (25.center) to (3.center);
		\draw [style=downArrow] (4.center) to (26.center);
		\draw [style=downArrow] (5.center) to (27.center);
		\draw [style=downArrow] (28.center) to (6.center);
		\draw [style=downArrow] (7.center) to (29.center);
		\draw (31.center) to (33);
		\draw (30.center) to (32);
		\draw [style=downArrow, bend left=90, looseness=1.25] (26.center) to (25.center);
		\draw [style=downArrow] (0.center) to (36);
		\draw [style=downArrow] (1.center) to (34);
		\draw [style=downArrow] (2.center) to (35);
		\draw (42.center) to (43);
		\draw [style=downArrow] (40.center) to (37.center);
		\draw [style=downArrow, in=90, out=-135] (36) to (44.center);
		\draw (56.center) to (57);
		\draw [style=downArrow] (34) to (50.center);
		\draw [style=downArrow] (55.center) to (52.center);
		\draw [style=downArrow, bend left=90] (62.center) to (55.center);
		\draw [style=downArrow] (51.center) to (63);
		\draw (74.center) to (75);
		\draw [style=downArrow] (72.center) to (69.center);
		\draw [style=downArrow] (70.center) to (73.center);
		\draw [style=downArrow] (67) to (68.center);
		\draw [style=downArrow, bend right=90, looseness=1.25] (78.center) to (72.center);
		\draw [style=downArrow] (79.center) to (28.center);
		\draw [style=downArrow, bend right=90, looseness=1.25] (80.center) to (79.center);
		\draw (86.center) to (87);
		\draw [style=downArrow] (84.center) to (81.center);
		\draw [style=downArrow] (82.center) to (85.center);
		\draw [style=downArrow] (27.center) to (90.center);
		\draw [style=downArrow, bend right=90, looseness=1.25] (90.center) to (84.center);
		\draw [style=downArrow] (85.center) to (91.center);
		\draw [style=downArrow] (91.center) to (92.center);
		\draw [style=downArrow] (38.center) to (62.center);
		\draw [style=downArrow] (35) to (94.center);
		\draw [style=downArrow, in=90, out=-105] (63) to (67);
		\draw [style=downArrow] (44.center) to (78.center);
		\draw [style=downArrow, bend right=90, looseness=1.25] (95.center) to (40.center);
		\draw [style=downArrow, in=90, out=-75, looseness=0.75] (36) to (95.center);
		\draw [style=downArrow, in=90, out=-75, looseness=0.50] (63) to (80.center);
	\end{pgfonlayer}
\end{tikzpicture}

%% file: JohnLikesHisCodeSloppy.tikz
\begin{tikzpicture}
	\begin{pgfonlayer}{nodelayer}
		\node [style=none] (0) at (-11.75, 3.25) {};
		\node [style=none] (1) at (-7.5, 3.25) {};
		\node [style=none] (2) at (-3.75, 3.25) {};
		\node [style=none] (3) at (-1.75, 3.25) {};
		\node [style=none] (4) at (0.5, 3.25) {};
		\node [style=none] (5) at (2.5, 3.25) {};
		\node [style=none] (6) at (5.75, 3.25) {};
		\node [style=none] (7) at (9.75, 3.25) {};
		\node [style=new style 0] (8) at (-11.75, 4) {John};
		\node [style=new style 0] (9) at (-7.5, 4) {likes};
		\node [style=new style 0] (10) at (-2.75, 4) {his};
		\node [style=new style 0] (11) at (0.5, 4) {code};
		\node [style=new style 0] (12) at (2.5, 4) {Bill};
		\node [style=new style 0] (13) at (7.75, 4) {does too};
		\node [style=none] (14) at (-11.175, 3) {$\F\semantics{N}$};
		\node [style=lab] (15) at (-6.75, 3) {$\F((\F(\F\semantics{N} \Rightarrow \semantics{S}))\Leftarrow \semantics{N})$};
		\node [style=lab] (16) at (-4.25, 1.25) {$\F(\F\semantics{N}\Rightarrow \semantics{N}) $};
		\node [style=none] (17) at (-1.175, 3) {$\semantics{N}$};
		\node [style=none] (18) at (1.075, 3) {$\semantics{N}$};
		\node [style=none] (19) at (3.075, 3) {$\F\semantics{N}$};
		\node [style=none] (20) at (7.05, 3) {$\F(\F\semantics{N} \Rightarrow \semantics{S})$};
		\node [style=none] (21) at (11.25, 3) {$\F\semantics{N} \Rightarrow \semantics{S}$};
		\node [style=none] (22) at (-1.75, 1.25) {};
		\node [style=none] (23) at (0.5, 1.25) {};
		\node [style=none] (25) at (5.75, 0.75) {};
		\node [style=none] (26) at (9.75, 0.5) {};
		\node [style=none] (27) at (5.75, 1.25) {};
		\node [style=none] (28) at (-1.75, 2.25) {};
		\node [style=new style 1] (29) at (9.75, 1.25) {};
		\node [style=new style 1] (30) at (-3.75, 2.25) {};
		\node [style=new style 0] (31) at (-7.5, -7) {$\Delta_{(\F(\F\semantics{N}\Rightarrow \semantics{S}))\Leftarrow \semantics{N}}$};
		\node [style=new style 0] (32) at (-3.75, 0) {$\Delta_{ \F\semantics{N} \Rightarrow \semantics{N} }$};
		\node [style=new style 0] (33) at (-11.75, 0.25) {$\Delta_{\semantics{N}}$};
		\node [style=none] (34) at (9.75, 0) {$\vdots$};
		\node [style=none] (35) at (-13.75, -2.25) {};
		\node [style=none] (36) at (-10, -2.25) {};
		\node [style=new style 0] (43) at (2.5, -0.25) {$\Delta_{\semantics{N}}$};
		\node [style=none] (44) at (9, -0.75) {};
		\node [style=none] (45) at (10.5, -0.75) {};
		\node [style=none] (46) at (9, -3) {};
		\node [style=none] (47) at (10.5, -2.5) {};
		\node [style=none] (48) at (9, -1.5) {};
		\node [style=new style 1] (49) at (10.5, -1.5) {};
		\node [style=none] (50) at (3.5, -3) {};
		\node [style=none] (51) at (10.5, -17.75) {};
		\node [style=none] (52) at (11.1, -1) {$\semantics{S}$};
		\node [style=none] (53) at (8.35, -1) {$\F\semantics{N}$};
		\node [style=none] (56) at (-2, -4) {};
		\node [style=new style 0] (57) at (1, -3.5) {$\epsilon_{\F\semantics{N}\Rightarrow \semantics{N}}$};
		\node [style=none] (58) at (1, -5.5) {};
		\node [style=none] (59) at (1, -5.75) {$\vdots$};
		\node [style=none] (62) at (0, -6.5) {};
		\node [style=none] (63) at (2, -6.5) {};
		\node [style=none] (64) at (0, -8.5) {};
		\node [style=none] (65) at (2, -8.5) {};
		\node [style=none] (66) at (0, -7.25) {};
		\node [style=new style 1] (67) at (2, -7.25) {};
		\node [style=none] (68) at (2.75, -6.75) {$\semantics{N}$};
		\node [style=none] (69) at (0.575, -6.75) {$\F\semantics{N}$};
		\node [style=none] (70) at (-2, -8.5) {};
		\node [style=new style 0] (71) at (-4.75, -2.5) {$\epsilon_{\F\semantics{N}\Rightarrow \semantics{N}}$};
		\node [style=none] (72) at (-4.75, -3.5) {};
		\node [style=none] (73) at (-4.75, -4) {$\vdots$};
		\node [style=none] (74) at (-5.5, -4.5) {};
		\node [style=none] (75) at (-4, -4.5) {};
		\node [style=none] (76) at (-5.5, -5) {};
		\node [style=none] (77) at (-4, -7.75) {};
		\node [style=none] (78) at (-5.5, -5) {};
		\node [style=new style 1] (79) at (-4, -5) {};
		\node [style=none] (80) at (-3.5, -4.5) {$\semantics{N}$};
		\node [style=none] (81) at (-6.15, -4.5) {$\F\semantics{N}$};
		\node [style=none] (82) at (-10, -5) {};
		\node [style=none] (83) at (-2, -11.75) {$\vdots$};
		\node [style=none] (84) at (-2.75, -12.25) {};
		\node [style=none] (85) at (-1.25, -12.25) {};
		\node [style=none] (86) at (-2.75, -13.5) {};
		\node [style=none] (87) at (-1.25, -13.5) {};
		\node [style=none] (88) at (-1.25, -13.5) {};
		\node [style=new style 1] (89) at (-2.75, -13.5) {};
		\node [style=new style 0] (90) at (-2, -10.5) {$\epsilon_{(\F(\F\semantics{N}\Rightarrow \semantics{S}))\Leftarrow \semantics{N}}$};
		\node [style=none] (91) at (-2, -11.5) {};
		\node [style=none] (92) at (2, -13.5) {};
		\node [style=none] (93) at (-0.75, -12.5) {$\semantics{N}$};
		\node [style=none] (94) at (-4.75, -14.5) {$\F(\F\semantics{N}\Rightarrow \semantics{S})$};
		\node [style=none] (95) at (5.75, -13.5) {};
		\node [style=none] (96) at (-8.5, -11) {$\vdots$};
		\node [style=none] (97) at (-9.5, -11.5) {};
		\node [style=none] (98) at (-7.5, -11.5) {};
		\node [style=none] (100) at (-7.5, -12.75) {};
		\node [style=none] (101) at (-7.5, -12) {};
		\node [style=new style 1] (102) at (-9.5, -12) {};
		\node [style=new style 0] (103) at (-8.5, -9.25) {$\epsilon_{(\F(\F\semantics{N}\Rightarrow \semantics{S}))\Leftarrow \semantics{N}}$};
		\node [style=none] (104) at (-8.5, -10.5) {};
		\node [style=none] (105) at (-7, -11.75) {$\semantics{N}$};
		\node [style=none] (106) at (-11.5, -12.25) {$\F(\F\semantics{N}\Rightarrow \semantics{S})$};
		\node [style=none] (107) at (-6.5, -12.75) {};
		\node [style=none] (108) at (-13.75, -16.5) {};
		\node [style=new style 0] (109) at (-9.5, -13) {$\epsilon_{\F\semantics{N} \Rightarrow \semantics{S}}$};
		\node [style=none] (110) at (-9.5, -14.25) {};
		\node [style=none] (111) at (-9.5, -14.5) {$\vdots$};
		\node [style=none] (112) at (-10.25, -15.25) {};
		\node [style=none] (113) at (-8.75, -15.25) {};
		\node [style=none] (114) at (-10.25, -16.5) {};
		\node [style=none] (115) at (-8.75, -17.75) {};
		\node [style=none] (116) at (-10.25, -15.75) {};
		\node [style=new style 1] (117) at (-8.75, -15.75) {};
		\node [style=none] (119) at (-11, -15.75) {$\F\semantics{N}$};
		\node [style=none] (120) at (-8.25, -17) {$\semantics{S}$};
		\node [style=none] (121) at (11, -17) {$\semantics{S}$};
		\node [style=none] (122) at (-13.2, -2.75) {$\F\semantics{N}$};
		\node [style=none] (123) at (4.075, -1.75) {$\F\semantics{N}$};
	\end{pgfonlayer}
	\begin{pgfonlayer}{edgelayer}
		\draw [style=downArrow] (22.center) to (3.center);
		\draw [style=downArrow] (4.center) to (23.center);
		\draw [style=downArrow] (25.center) to (6.center);
		\draw [style=downArrow] (7.center) to (26.center);
		\draw (28.center) to (30);
		\draw (27.center) to (29);
		\draw [style=downArrow, bend left=90, looseness=1.25] (23.center) to (22.center);
		\draw [style=downArrow] (0.center) to (33);
		\draw [style=downArrow] (1.center) to (31);
		\draw [style=downArrow] (2.center) to (32);
		\draw [style=downArrow, in=90, out=-150] (33) to (35.center);
		\draw [style=downArrow, in=90, out=-30] (33) to (36.center);
		\draw [style=downArrow] (5.center) to (43);
		\draw [style=downArrow, bend right=90, looseness=1.50] (50.center) to (46.center);
		\draw [style=downArrow] (46.center) to (44.center);
		\draw [style=downArrow] (45.center) to (47.center);
		\draw (48.center) to (49);
		\draw [style=downArrow] (47.center) to (51.center);
		\draw [style=downArrow] (57) to (58.center);
		\draw [style=downArrow] (64.center) to (62.center);
		\draw [style=downArrow] (63.center) to (65.center);
		\draw (66.center) to (67);
		\draw [style=downArrow] (56.center) to (70.center);
		\draw [style=downArrow, bend right=90, looseness=1.50] (70.center) to (64.center);
		\draw [style=downArrow] (71) to (72.center);
		\draw [style=downArrow] (76.center) to (74.center);
		\draw [style=downArrow] (75.center) to (77.center);
		\draw (78.center) to (79);
		\draw [style=downArrow] (36.center) to (82.center);
		\draw [style=downArrow, bend right=90, looseness=0.75] (82.center) to (76.center);
		\draw [style=downArrow] (90) to (91.center);
		\draw [style=downArrow] (84.center) to (86.center);
		\draw [style=downArrow] (87.center) to (85.center);
		\draw (88.center) to (89);
		\draw [style=downArrow] (65.center) to (92.center);
		\draw [style=downArrow, bend left=90, looseness=1.50] (92.center) to (87.center);
		\draw [style=downArrow, bend right=90, looseness=1.25] (86.center) to (95.center);
		\draw [style=downArrow] (95.center) to (25.center);
		\draw [style=downArrow] (103) to (104.center);
		\draw [style=downArrow] (100.center) to (98.center);
		\draw (101.center) to (102);
		\draw [style=downArrow, in=90, out=-90, looseness=1.25] (77.center) to (107.center);
		\draw [style=downArrow, bend left=90, looseness=1.50] (107.center) to (100.center);
		\draw [style=downArrow] (35.center) to (108.center);
		\draw [style=downArrow] (97.center) to (109);
		\draw [style=downArrow] (109) to (110.center);
		\draw [style=downArrow] (114.center) to (112.center);
		\draw [style=downArrow] (113.center) to (115.center);
		\draw (116.center) to (117);
		\draw [style=downArrow, bend right=90, looseness=1.25] (108.center) to (114.center);
		\draw [style=downArrow, in=90, out=-135] (31) to (103);
		\draw [style=downArrow, in=90, out=-135, looseness=1.25] (32) to (71);
		\draw [style=downArrow, in=90, out=-30, looseness=0.75] (32) to (57);
		\draw [style=downArrow, in=90, out=-30, looseness=0.75] (31) to (90);
		\draw [style=downArrow, in=90, out=-135] (43) to (56.center);
		\draw [style=downArrow, in=90, out=-45] (43) to (50.center);
	\end{pgfonlayer}
\end{tikzpicture}